\documentstyle[12pt,aasms4]{article}

\slugcomment{Accepted for Publication in \it The Astrophysical Journal\rm}

\lefthead{Gibson et~al.}
\righthead{HST Key Project: Type Ia SNe and H$_\circ$}

\begin{document}

\title{The $HST$ Key Project on the Extragalactic Distance Scale
XXV.  A Recalibration of Cepheid Distances to Type Ia
Supernovae and the Value of the Hubble Constant\footnote{Based on 
observations with the 
NASA/ESA \it Hubble Space Telescope\rm, obtained at the Space Telescope 
Science Institute, which is operated by AURA, Inc., under NASA Contract No. 
NAS 5-26555.}}

\author{Brad K. Gibson\altaffilmark{2}, 
Peter B. Stetson\altaffilmark{3}, 
Wendy L. Freedman\altaffilmark{4}, 
Jeremy R. Mould\altaffilmark{5},
Robert C. Kennicutt, Jr.\altaffilmark{6},
John P. Huchra\altaffilmark{7}, 
Shoko Sakai\altaffilmark{8},
John A. Graham\altaffilmark{9}, 
Caleb I. Fassett\altaffilmark{10},
Daniel D. Kelson\altaffilmark{9},
Laura Ferrarese\altaffilmark{11}, 
Shaun M.G. Hughes\altaffilmark{12},
Garth D. Illingworth\altaffilmark{13}, 
Lucas M. Macri\altaffilmark{7},
Barry F. Madore\altaffilmark{14},
Kim M. Sebo\altaffilmark{5} and
Nancy A. Silbermann\altaffilmark{14}}

\altaffiltext{2}{Center for Astrophysics \& Space Astronomy, Department of Astrophysical \& Planetary Sciences, University of Colorado, Campus Box 389, Boulder, CO, USA  80309-0389}
\altaffiltext{3}{Dominion Astrophysical Observatory, Herzberg Institute of Astrophysics, National Research Council, 5071 West Saanich Rd., Victoria, BC, Canada  V8X~4M6} 
\altaffiltext{4}{The Observatories, Carnegie Institution of Washington, Pasadena, CA, USA  91101}
\altaffiltext{5}{Research School of Astronomy \& Astrophysics, Australian National University, Weston Creek Post Office, Weston, ACT, Australia  2611}
\altaffiltext{6}{Steward Observatory, University of Arizona, Tucson, AZ, USA  85721}
\altaffiltext{7}{Harvard College, Center for Astrophysics, 60 Garden St., Cambridge, MA, USA  02138}
\altaffiltext{8}{National Optical Astronomy Observatories, P.O. Box 26732, Tucson, AZ, USA  85726}
\altaffiltext{9}{Department of Terrestrial Magnetism, Carnegie Institution of Washington, 5241 Broad Branch Rd. N.W., Washington, D.C., USA  20015}
\altaffiltext{10}{Department of Astronomy, Williams College, Williamstown, MA, USA  01267}
\altaffiltext{11}{Palomar Observatory, California Institute of Technology, Pasadena, CA, USA  91125} 
\altaffiltext{12}{Institute of Astronomy, Madingley Road., Cambridge, UK  CB3~0HA}
\altaffiltext{13}{Lick Observatory, University of California, Santa Cruz, CA, USA 95064}
\altaffiltext{14}{Infrared Processing and Analysis Center, Jet Propulsion Laboratory, California Institute of Technology, Pasadena, CA, USA  91125}

\def\spose#1{\hbox to 0pt{#1\hss}}
\def\simlt{\mathrel{\spose{\lower 3pt\hbox{$\mathchar"218$}}
     \raise 2.0pt\hbox{$\mathchar"13C$}}}
\def\simgt{\mathrel{\spose{\lower 3pt\hbox{$\mathchar"218$}}
     \raise 2.0pt\hbox{$\mathchar"13E$}}}
\def\eg{{\rm e.g.}}
\def\ie{{\rm i.e.}}
\def\etal{{\rm et~al.}}
\def\ho{{\rm H$_\circ$}}
\def\h0{{\rm H$_\circ$}}
\def\hounits{{\rm km\,s$^{-1}$\,Mpc$^{-1}$}}
\def\tf{{\rm Tully-Fisher }}
\def\ct{{\rm Cal\'{a}n-Tololo }}
\def\bmv{\hbox{\it B--V\/}}
\def\vmi{\hbox{\it V--I\/}}

\begin{abstract}
Cepheid-based distances to seven Type Ia supernovae (SNe)-host galaxies have
been derived using the standard \it HST Key Project on the Extragalactic
Distance Scale \rm pipeline.  For the first time, this allows for a transparent
comparison of data accumulated as part of three different \it HST \rm projects,
the \it Key Project\rm, the Sandage \etal\ Type Ia SNe program, and the Tanvir
\etal\ Leo~I Group study.  Re-analyzing the Tanvir \etal\ galaxy and
six Sandage \etal\ galaxies we find a mean (weighted)
offset in true distance moduli
of $0.12\pm 0.07$\,mag -- \ie, 6\% in linear distance -- in the sense of 
reducing
the distance scale, or increasing H$_\circ$.  Adopting the
reddening-corrected Hubble relations of Suntzeff \etal\ (1999), tied to a
zero point based upon SNe~1990N, 1981B, 
1998bu, 1989B, 1972E and 1960F and the photometric calibration of 
Hill \etal\ (1998), leads to a
Hubble constant of H$_\circ=68\pm 2\,({\rm random})\pm 5\,({\rm 
systematic})$ \hounits.  
Adopting the Kennicutt \etal\ (1998) Cepheid period-luminosity-metallicity
dependency decreases the inferred H$_\circ$ by 4\%.
The H$_\circ$ result from Type Ia SNe is now in good agreement, to
within their respective uncertainties, with that from the Tully-Fisher
and surface brightness fluctuation relations.
\end{abstract}

\keywords{Cepheids --- distance scale --- galaxies: distances and redshifts 
--- supernovae: general}

\section{Introduction}
\label{intro}

Type Ia supernovae (SNe) have recently been used extensively
for the measurement of relative distances to galaxies
(\eg, Hamuy \etal\ 1996;
Filippenko \& Riess 1999; Branch 1998; and
Riess, Press \& Kirshner 1996).
With peak $B$-band absolute magnitudes ranging (typically) from
$<{\rm M}^{\rm max}_B> \approx -18.8$ to $-19.8$, Type Ia SNe have been
discovered in ever-increasing numbers at
redshifts ($z \approx 0.01 \rightarrow 0.1$) well out in the Hubble 
Flow -- the 29 Type Ia SNe of the \ct Survey (Hamuy \etal\ 1996),
supplemented by the seven new SNe from the Center for Astrophysics (CfA)
Survey (Riess \etal\ 1998), are
testament to this fact.  The dispersion in the  raw \ct
Survey $V$-band Hubble diagram alone is only $\sigma = 0.26$, $\sim$0.15\,mag
tighter than the $V$-band \tf relationship (Sakai \etal\ 
1999).  

In recent years,  it has become clear that Type Ia SNe 
are not standard candles of constant luminosity.
Significant dispersion in 
their peak luminosities, particularly in the $B$ and $V$ bands, has now been
unequivocally demonstrated (\eg, Phillips 1993; Hamuy \etal\ 1996; Riess \etal\ 
1996, 1998; Phillips \etal\ 1999); fortuitously, though, these same groups
emphasize that there is a strong correlation between SNe light-curve shape and
peak luminosity.  This is perhaps best appreciated through Figure 8 of Phillips
\etal\ (1999), in which the
correlation between peak luminosity and decline rate
$\Delta m_{\rm 15}(B)$ is shown. $\Delta m_{\rm 15}(B)$ corresponds
to the decline in $B$-band magnitude from maximum to $t=15$ days after $B$
maximum.
Correcting the observed peak magnitudes for this decline rate effect results in
a $V$-band Hubble diagram dispersion of 0.14\,mag (Hamuy \etal\ 1996).

While the Hubble diagrams, whether they be raw or corrected for
decline-rate effects,
can provide highly accurate relative distances, the
derivation of an actual Hubble constant, H$_\circ$,
requires that a locally-calibrated 
zero point be provided; with the launch of $HST$, this zero point
can now be provided via independent,
Cepheid-based, distance
determinations for nearby Type Ia SNe-host galaxies.
Providing such a zero point is the goal of 
the Sandage/Saha {\it Type Ia SNe HST Calibration Program} (Saha \etal\ 1997, 
and references therein).
To date, six galaxies have been targeted by the Sandage/Saha program 
(NGC~3627, 4496A, 4536, 4639, 5253, and IC 4182); each galaxy's associated 
Type Ia SN, metallicity, and {\it HST} Proposal ID number is listed in Table 
\ref{tbl:galaxies}.  
This list of six is supplemented by two others -- NGC 3368 (host to 
SN~1998bu), from Tanvir \etal\ (1995) and NGC 4414 (host to SN 1974G), from 
Turner \etal\ (1998).  Three further local Type Ia-host galaxies 
will be observed by Sandage/Saha during {\it HST} Cycles 7--9 
(NGC~1316, 4527, and 3982) and, for 
completeness, are also listed in Table \ref{tbl:galaxies}.

\placetable{tbl:galaxies}

As part of the
\it HST Key Project on the Extragalactic Distance Scale\rm, Cepheid-based
distance determinations have been made for 18 galaxies, including one
local, calibrating, Type Ia SNe-host (NGC~4414 -- Turner \etal\ 1998).  This
ensemble is now being employed as calibration for a variety of secondary
distance indicators probing further into the Hubble flow, including the \tf
relationship (Sakai \etal\ 1999), surface brightness fluctuations (Ferrarese
\etal\ 1999), fundamental plane (Kelson \etal\ 1999), and, of course, Type Ia 
SNe (this work).  

This paper differs somewhat from these other calibrations
of secondary methods in that we now focus on the recalibration
of extant Cepheid data. It presents results from a search
for Cepheids in all galaxies observed with \it HST \rm that are host to Type
Ia supernovae.
The goal of the present analysis is to process the remaining non-\it Key
Project\rm, \it HST\rm-observed, Type Ia SNe-host galaxies in a manner
consistent with that employed throughout the earlier papers in our series.
While parallel double-blind reductions were not employed for the present
recalibration work (as was done for Papers I-XXIII), the philosophy adopted for
the seven galaxies described herein \it is \rm consistent with that used for
our earlier \it Key Project \rm galaxies.  This
ensures that these galaxies are on the same footing as our original sample of
18, and provides a useful cross-check of the Sandage/Saha and Tanvir \etal\ 
independent reductions.  A
benefit of our re-analysis is that not only do we have a better understanding
of the subtleties of accurately calibrating WF/PC and WFPC2 data, but we are 
in a position to take advantage of
additional archive data for one of the galaxies (NGC~5253), 
that was not available to the original authors.

In Section \ref{previous}, results from previous Type Ia supernova
studies are briefly summarized.
In Section \ref{data}, 
we briefly remind the reader of this standard {\it HST Key 
Project} reduction and calibration pipeline,
then discuss the new photometric reductions, Cepheid
period-luminosity (PL) relations, and distances to each Cepheid
calibrator galaxy,  and then compare them
to previous results.
The Cepheid photometry for our reanalysis is provided in an appendix. 
Section \ref{implications}
is devoted to the 
implications for H$_\circ$ work implied by our uniform re-analysis of the local 
Type Ia SNe calibrators; these implications hold for  all the empirical 
studies of Table \ref{tbl:H0_Ia}; 
for brevity, only two  analyses (those due to 
Suntzeff \etal\ 1999 and Saha \etal\ 1997) are followed in detail.  The role 
of a possible metallicity correction to the Cepheid-derived distances 
is discussed in Section \ref{metallicity}, while the local flow field and its 
impact upon H$_\circ$ is described in Section \ref{flow}.  We 
re-visit the Type Ia SNe-predicted distance to the Fornax Cluster
in Section \ref{fornax}.
Finally, our results are summarized in Section \ref{summary}.

\section{Previous Results}
\label{previous}

Using their
five Cepheid-based distances to local Type Ia SNe-host galaxies, 
supplemented with an indirectly inferred distance to NGC~3627, Saha \etal\ 
(1997) derived zero points to their  uncorrected
$B$- and $V$-band Hubble diagrams (themselves, based
equally upon the \ct and Asiago (Patat 1995) surveys).  Ignoring external
systematic effects, their resulting Hubble constant was \ho$ =58 \pm 3$
\hounits.  Using only the uncorrected \ct sample, but now including
the $I$-band data, and ignoring those calibrators with poor photometry or
without a direct Cepheid-based distance determination, both Hamuy \etal\ (1996)
and Riess \etal\ (1996) found \ho$=56 \pm 2$ \hounits, in agreement with Saha
et~al.

Correcting \rm for the light-curve
shape-peak luminosity relationship, in both the high-$z$ \it and \rm local
calibrator sample), Hamuy \etal\ (1996), Riess \etal\ 
(1996,1998), Phillips \etal\ (1999), and Suntzeff \etal\ (1999), each
found \ho$=64 \pm 3$ \hounits, despite the different approaches
in the treatment of foreground and host galaxy extinction -- \ie, correcting the
peak magnitudes for light-curve shape results in an approximate
$10 - 15$\% increase in \ho.

A variety of other Type Ia SNe-related H$_\circ$ determinations populate the
literature. The
subtle differences in each study are tied to the details of either the
high-$z$ or local calibrator sample selection.  For example, 
a preliminary analysis from our team (Madore \etal\ 1999) led to a value for
the Hubble constant of 
H$_\circ$=$67\pm 6\,({\rm random})\pm 7\,({\rm systematic})$.  For that study
though, only three of the six Type Ia SNe employed had direct
Cepheid-calibrated distances; indirect distances to two Fornax Cluster SNe
were inferred via the Cepheid distance to cluster member NGC~1365 (Silbermann
\etal\ 1999; see also Section \ref{fornax}).  
In the analysis which follows we have chosen to restrict
ourselves only to those eight SNe which have \it direct \rm Cepheid-calibrated
distances; as Madore \etal\ rightly note, the inclusion of the two Fornax SNe
has negligible impact upon the derived H$_\circ$.

An alternative method, independent of the local, empirical
calibration  of the Hubble constant, is the derivation of 
H$_\circ$ based upon purely physical models of Type Ia SNe
explosions.
This method will not be discussed in any
detail in what follows, except to note that it is encouraging that the weighted mean 
of three recent {\it theoretical} Type Ia SNe-H$_\circ$
analyses (H\"{o}flich \& 
Khokhlov 1996; Ruiz-Lapuente 1996; Iwamoto \& Nomoto 1998) is \ho$= 
67 \pm 6$ \hounits, in close agreement with the
recent empirical values, to be discussed in Section \ref{hamuy}.

Table \ref{tbl:H0_Ia}
provides a useful summary of these recent Type Ia SNe-based H$_\circ$
determinations,
including both physical and empirical methodologies.  Where available (and 
where relevant), an attempt has been made to list the number of SNe included
in the adopted Hubble diagrams, as well as the number of local calibrators 
employed in deriving the associated zero points.  The associated mean peak 
$B$-band magnitude for each sample is likewise provided; the notes to Table 
\ref{tbl:H0_Ia}
clarify whether said mean applies to the high-$z$ sample or the local 
calibrators.

\placetable{tbl:H0_Ia}

\section{The Data}
\label{data}

Seven Type Ia SNe-host galaxies, six from the Sandage/Saha team (NGC~4496A,
4536, 4639, 5253, 3627 and IC~4182) and one from Tanvir \etal\ (NGC~3368), were
extracted from the \it HST \rm archive and processed with the \it Key
Project\rm's ALLFRAME (photometry -- Stetson 1994) and TRIAL (variable
finding -- Stetson 1996) pipeline, as documented thoroughly by Stetson \etal\
(1998) and Hill \etal\ (1998).  For Papers I-XXIII of the \it Key Project\rm,
we carried out a (real-time) parallel reduction,  based on both DoPHOT and
ALLFRAME, using at least two different methods for finding the
Cepheids.  We included in the final analysis only those Cepheids which were
in common to both analyses, but for consistency, generally quoted only
the ALLFRAME results.  In the present  case, the DoPHOT results were
already previously published by the Sandage/Saha team -- \ie, we did not 
undertake a parallel DoPHOT reduction and, thus, the methodology employed is
not \it identical \rm to that used in Papers I-XXIII of this series.

For the ALLFRAME$+$TRIAL analysis described here, photometry, Cepheid
selection, and PL analyses were handled identical to each of our previous
galaxies -- \ie, 
no special treatment
was afforded any of the Cepheid samples herein, and only those Cepheids with
good quality $V$- \it and \rm $I$-band data were considered.  As described in
Sections \ref{n4639}-\ref{n4496a}, for several of the galaxies analyzed by Saha
et~al., these authors either (a) used a special PL-fitting scheme 
(\eg, NGC~4639 and 3627), or (b) treated the $I$-band data specially in the
derivation of the true modulus (\eg, NGC~4496A and IC~4182).
For those galaxies with peculiarities not
encountered in earlier \it Key Project \rm papers,
details are provided in the appropriate subsections below.  
We discuss the reasons for the differences between previously published 
results and those in this paper, so that the reasons for the
difference in the overall value of H$_0$ derived in this paper are clear.
As it turns out, there is not a single explanation for the differences
we see compared to our standard pipeline processing procedure, but a variety of reasons.
The full set of
Cepheids for each of the seven galaxies, is listed in Appendix A, along with
positional and photometric information.  As alluded to previously, we have only
retained those Cepheids which have high quality $V$- \it and \rm $I$-band
photometry; we made a conscious decision to not retain the lower quality
candidates.
Because of this, the Cepheid numbers in Tables A1-A7 are
typically smaller than 
the number listed in the corresponding Saha \etal\ DoPHOT
analysis, since Saha \etal\ provided a complete listing of both
bona fide \it and
\rm lower-quality suspected Cepheid candidates.
Epoch-by-epoch photometry,
light-curves, and finding charts, for all the Cepheids discovered in this work
are available through the \it HST Key Project\rm's web 
page.\footnote{\tt http://www.ipac.caltech.edu/H0kp/}

The \it Key Project\rm's 
data were calibrated without the adoption of an explicit, analytic,
correction for charge transfer
effects (CTE), although the Hill \etal\ (1998) calibration employed did correct
for it in the mean.  On the other hand, the 
current data \it were \rm analyzed with a developmental
calibration which does treat CTE (Stetson 1998), although our quoted 
moduli, and the subsequent H$_\circ$ analysis which follows, 
adopt the Mould \etal\ (1999b) transformation from the Stetson (1998; S98) 
scale to that of Hill \etal\ (1998; H98); 
this amounts to mean shifts of $+$0.02\,mag ($V$) and
$+$0.04\,mag ($I$), in the sense of H98 being greater than S98.  This
corresponds to a $+$0.07\,mag shift (S98$\rightarrow$H98)
in the true modulus, and retains
consistency with previous (and current)
papers in this series.  The two systems do agree at the $\sim$1$\sigma$ level,
but unresolved discrepancies between the two have led us to conservatively
adopt the Hill \etal\ (1998) scale, in what follows.

\subsection{NGC~4639}
\label{n4639}

NGC~4639, host to SN~1990N, is one of three local calibrators situated 
in the Virgo Cluster.
Of the 38 
galaxies (to date) with Cepheid-based distances (Ferrarese \etal\ 1999; 
Table 3), NGC~4639 is the current record-holder; Saha \etal\ (1997) 
derive a true modulus of $\mu_\circ=32.03\pm 0.22$, based
upon 17 Cepheid candidates.  
Our ALLFRAME$+$TRIAL analysis
uncovered 17 high-quality Cepheid candidates, 
seven of which were in common with the Saha \etal\ (1997; Table~4) sample.  
The majority of our sample (\ie, 14/17) reside on Chips~2 and 3.  
Positional and photometric information for the full sample is given in 
Table \ref{tbl:n4639_ceph}.  The first entry of Table \ref{tbl:ceph_comp}
shows that the agreement in
assigned mean $V$-band magnitudes, for the seven Cepheids in common between 
Table
\ref{tbl:n4639_ceph} and Saha \etal\ (Table~4), is excellent.  
However, as is the case for many of the galaxies discussed in this
paper, the agreement in the
$I$-band is somewhat inferior (although the agreement does hold at the
$\sim 1\sigma$ level).

\placetable{tbl:ceph_comp}

Our $V$- and $I$-band PL relations are
shown in Figure \ref{fig:n4639}.  
Standard PL-fitting to the full sample
results in $\mu_V=31.92\pm 0.08$, $\mu_I=31.87\pm 0.06$, 
E$(\vmi)=0.05\pm 0.03$, and $\mu_\circ=31.80\pm 0.07$; the quoted
uncertainties simply reflect the rms scatter of the fits, which includes the
effects of the finite width of the instability strip as well as of random
observational noise.  A full accounting of
the error budget (detailed in Table \ref{tbl:error})
leads to our final adopted
value of $\mu_\circ=31.80\pm 0.09\,({\rm random})\pm 0.16\,({\rm systematic})$.
In comparison, recall that Saha \etal\ (1997) found
$\mu_\circ=32.03\pm 0.22$ -- \ie, our result places NGC~4639 
0.2\,mag nearer than that favored by Saha et~al (although still in agreement
at the $1\sigma$ level).

\placefigure{fig:n4639}

Saha \etal\ (1997) interpret their DoPHOT PL
relations as showing evidence for significant scatter beyond that encountered
in previous galaxies (indicative of a substantial component of
differential reddening), and thereafter employ a weighting scheme in averaging
the sample of de-reddened Cepheid moduli, in order to derive the 
galaxy's true modulus.  We would suggest that upon closer inspection
the evidence presented 
therein for differential reddening is perhaps not as suggestive as might at
first seem.  

First, of the 15 Cepheids employed by Saha \etal\ (neglecting the
same three which they discard in deriving the true modulus, 
but which are still plotted in their apparent $V$- and $I$-band moduli PL
relations - \ie, their C1-V3, C3-V8, and C3-V9), only two lie outside the LMC
$V$-band $2\sigma$ instability strip width ($\pm 0.54$\,mag).  Our analysis
(upper panel of Figure \ref{fig:n4639}) found one candidate lying outside this
region.  Formally, the random uncertainties 
associated with the dereddened PL fit
to the respective samples (\ie, entry R3 of Table \ref{tbl:error}) are $\pm
0.07$\,mag (ALLFRAME) and $\pm 0.10$\,mag (DoPHOT), so while the scatter
in the Saha \etal\ dereddened moduli distribution is marginally larger than
that found in our ALLFRAME$+$TRIAL analysis, in neither case is the scatter
anomalously large in comparison with the entire set of \it HST\rm-observed
galaxies.  A simple visual inspection of
our ALLFRAME PL relations (and the listed $V$- and $I$-band scatter in each
panel) for NGC~4639 (Figure \ref{fig:n4639}), 
in comparison with those for the other galaxies in this study (Figures
\ref{fig:n4536}-\ref{fig:n4496a}),
do not show a significantly larger scatter.  One can also do a formal fit to the
distribution of Cepheids in the differential reddening diagnostic plane (\ie,
$\mu_V-\mu_I$ versus $\mu_V$ - 
Figure 7 of Saha \etal\ 1997) -- for both the ALLFRAME$+$TRIAL and DoPHOT
datasets, the formal slope of the least squares fit is $0.3\pm 0.1$.  A slope
of $0.20$ is expected in the case of photometric scatter alone, while a slope
of $0.41$ is expected in the case where differential reddening is substantial
- \ie, statistically, we feel there is no compelling reason to invoke any
substantial differential reddening component, and for these reasons, we
have treated NGC~4639 in the same way as all of the other galaxies
we have analyzed, taking the unweighted mean as the representative final galaxy
modulus.  The differing weighting schemes employed account for half of the
discrepancy between the two analyses, while the remainder is
due to the aforementioned (unresolved) $I$-band disagreement.

\subsection{NGC~4536}
\label{n4536}

NGC~4536 was host to SN~1981B, for which the light curve 
and photometry are of the 
highest quality.  Saha \etal\ (1996a) derived moduli of $\mu_V=31.23\pm 
0.05$, $\mu_I=31.18\pm 0.05$, and 
$\mu_\circ=31.10\pm 0.13$, and a total 
line-of-sight reddening of E$(\vmi)=0.05\pm 0.04$, based upon the 27 
highest-quality Cepheids (\ie, $\rm{QI}\ge 5$) in their Table~3.
Our ALLFRAME$+$TRIAL analysis resulted in the sample of 39 candidates listed 
in Table \ref{tbl:n4536_ceph}.  The $V$- and $I$-band PL relations
are shown in Figure \ref{fig:n4536}; 
because of incompleteness below periods of 20 days,
the formal PL fits were restricted to the 27 Cepheids 
for which $P>20\,$d, and are represented by the solid lines
(\ie, $\mu_V=31.20\pm 0.06$ and $\mu_I=31.10\pm 0.04$).  
Our final true modulus is $\mu_\circ=30.95\pm 0.07\,({\rm
random})\pm 0.16\,({\rm random})$.

\placefigure{fig:n4536}

Our predicted true distance modulus
appears 0.15\,mag
discrepant with that of Saha \etal\ (1996a) ($\mu_\circ=31.10\pm 0.13$).
Examining the Saha \etal\ data on a chip-by-chip basis, demonstrates though
that this discrepancy results (partially)
from what appears to be a discrepant chip within their dataset.
If we apply the same color ($0.6\le {\rm E}(V-I)\le 1.4$), 
period (P$>$20\,d), and quality-index (QI$\ge 4$) cuts adopted by Saha 
\etal\ (1996a), and then perform the standard PL fitting to (a) the sum of Chips
$1+3+4$ (17 Cepheids), and (b) Chip 2 alone (12 Cepheids), we find that (a)
$\mu_\circ(1+3+4)=30.76\pm 0.12$\,(random) and (b) $\mu_\circ(2)=31.42\pm 
0.12$\,(random).  The upper panel of 
Figure \ref{fig:n4536_mod_comp} illustrates this 0.66\,mag Chip~2
discrepancy present within the Saha \etal\ dataset, where the solid symbols
represent the 17 Chip~1+3+4 Cepheids and the open symbols the 12 Chip~2
Cepheids; the solid and dotted horizontal lines correspond to the unweighted
mean of each respective sample.  The lower panel of Figure
\ref{fig:n4536_mod_comp} shows that such a discrepancy does not exist in our
ALLFRAME$+$TRIAL dataset.  Because the Saha \etal\ Chips 1, 3, and 4, are
self-consistent, we feel that the true modulus derived from the mean of the
Cepheids therein -- $\mu_\circ=30.76\pm 0.12\,({\rm random})\pm 0.16\,({\rm
systematic})$ -- represents a better option than the originally published
$\mu_\circ=31.10\pm 0.13$, as reflected by its entry under $\mu_\circ({\rm
KP})$ in Table \ref{tbl:distances}.  The fact that our ALLFRAME$+$TRIAL
distance to NGC~4536 is 0.19\,mag greater than that derived from Chips 1+3+4 of
the Saha \etal\ sample is consistent with that expected, based upon a
comparison of mean magnitudes for the
Cepheids in common in the two datasets
(see relevant entry in Table \ref{tbl:ceph_comp}).

\placefigure{fig:n4536_mod_comp}

\subsection{NGC~3627}
\label{n3627}

Host to the highly-reddened, but well-photometered,
SN~1989B (Wells \etal\ 1994), NGC~3627 is the most recently observed galaxy by
the Sandage/Saha Type Ia SNe Calibration Team.  The WFPC2 field is replete with
dust lanes, and suffers from an extremely bright background.  Based upon a
sub-sample of 25 Cepheids, drawn from a master list of 68, Saha \etal\ (1999)
derive an true modulus of $\mu_\circ=30.22\pm 0.12$.  

The ALLFRAME$+$TRIAL analysis for NGC~3627 proved to be more challenging than
that encountered for any previous galaxy.  The problem stems, in part, from a
distinct lack of bright local standards upon which to base the calibration.
There  are many bright objects in the field, but they are
decidedly non-stellar, forcing the use of those
comparatively faint calibrators which successfully avoid the dust lanes.
In addition, even after eliminating all
likely star clusters and background galaxies, the remaining stellar images are
still $\sim$50\% broader than in the rest of the \it HST-\rm observed galaxies.
In practice, what this means, is that the median aperture correction (in going
from 2 to 5 pixel radius), for NGC~3627, is $-$0.362\,mag, while the median for
all the other galaxies is $-$0.192\,mag --
\ie, there is a 0.17\,mag difference in
the aperture corrections for NGC~3627.

The source of this discrepancy is not obvious. 
Perhaps all of the NGC~3627 observations were out of 
focus; however, contemporaneous \it Key Project \rm WFPC2 
NGC~3319 observations do
not deviate from the median aperture correction noted previously.
Second, perhaps NGC~3627 is  extremely  rich in star clusters
which are incorrectly being identified as stars.
A third possibility is that the extremely high surface brightness of NGC~3627 
is problematic;
however, the form of the
stellar images in the low surface brightness regions of our
frames do not appear appreciably different from those in the high
surface brightness parts of the same images.

In what follows, we have chosen the conservative approach, and
adopted the second option - \ie, using the full sample's
median aperture corrections and making use of all of the
data available to us.  Instead of the standard 0.05\,mag random
photometry uncertainty used for other galaxies in this study
(\ie, R1 in Table \ref{tbl:error}), for NGC~3627 we have
assumed the more generous value of 0.15\,mag, to reflect the greater
uncertainty associated with the adopted aperture corrections.  Our results
should be reliable at this level.

In total, 36 high-quality
Cepheid candidates were discovered, the properties for which are listed in
Table \ref{tbl:n3627_ceph}.  Eighteen of our Cepheids are in common with the
Saha \etal\ (1999; Table 4) sample, and as can be seen by the relevant entry to
Table \ref{tbl:ceph_comp}, the agreement in the mean $V$-band
magnitudes is excellent, although the $I$-band agreement is somewhat inferior
(agreement at the 2$\sigma$ level only).  
Based upon this common sample, we \it a priori
\rm expect that our ALLFRAME true modulus should be 0.17\,mag \it greater
\rm than that of Saha \etal\ (1999).

The PL relations for NGC~3627 are shown in Figure
\ref{fig:n3627}, and possibly provide evidence for the presence of
differential reddening, by the excess scatter beyond that expected from the
finite width of the
LMC instability strip.  Larger errors as a result of the difficulty
in working against the high background may also contribute to the
higher scatter.
Incompleteness, for P$<$25\,d, is apparent,
so our preferred ALLFRAME$+$TRIAL distance has been based upon the 17 Cepheids
with periods in excess of 25\,d.  Standard PL fitting,
incorporating the full error budget, yields
$\mu_\circ=30.06\pm 0.17\,({\rm random})\pm 0.16\,({\rm systematic})$.  Had
we included Cepheids with periods less than 25\,days, our derived distance to
NGC~3627 would decrease.

\placefigure{fig:n3627}

In contrast, Saha \etal\ (1999) derive the significantly larger modulus
of $\mu_\circ=30.22\pm 0.12$.  This 0.16\,mag \it greater \rm modulus is at
odds with that which we had predicted on the basis of the 18 Cepheids in common
(recall, we expected the DoPHOT modulus to be 0.17\,mag \it smaller \rm than
our ALLFRAME modulus).  The source of the discrepancy appears to be due to the
sample of Cepheids used by both groups which are \it not \rm 
part of the aforementioned ``common'' subset.  Restricting ourselves to only
those Cepheids used by each group in their respective PL fitting (\ie, the 25
used by Saha \etal\ and the 17 used in our re-analysis), we show in Figure
\ref{fig:n3627_mod_comp} the distribution of individually dereddened true
moduli for each sample.  

\placefigure{fig:n3627_mod_comp}

The upper panel of Figure \ref{fig:n3627_mod_comp} shows the 25 Saha \etal\
(1999) Cepheids, while the lower panel shows our 17; in both panels,
the filled symbols represent the five Cepheids which occur in both samples.
The horizontal lines represent the unweighted mean of the dereddened moduli,
hence the $\mu_\circ=30.29$ shown in the upper panel, which differs from the
aforementioned Saha \etal\ value of $\mu_\circ=30.22$ (which represents their
favored weighted mean).  

Several points should be made regarding Figure \ref{fig:n3627_mod_comp} --
first, the mean magnitudes of the (small) overlap
sample are in good agreement, as was found for the larger sample described
earlier.  Indeed, just using these five Cepheids in common leads to
$\mu_\circ=30.17\pm 0.17$\,(random), \it 
regardless of whose photometry is adopted\rm.  Of the 0.23\,mag discrepancy
noted in Figure \ref{fig:n3627_mod_comp}, 0.14\,mag is being driven by the four
Saha \etal\ Cepheids with $\mu_\circ>30.8$; removal of these four outliers
leads to $\mu_\circ=30.15\pm 0.08$\,(random), in better agreement with our
favored $\mu_\circ=30.06\pm 0.17$\,(random).  Further support for their
excision comes from the fact that their inferred reddenings are all negative,
as witnessed by their populating the lower-right quadrant of Saha et~al.'s
Figure~11.  The residual 0.09\,mag offset can
be traced primarily 
to those $18<{\rm P}<28$\,d Cepheids in the Saha \etal\ sample which
were not recovered in our analysis; those Saha \etal\ Cepheids
\it not \rm included in our analysis (\ie, the 20 open squares in the upper
panel of Figure \ref{fig:n3627_mod_comp}), appear to lie, in the mean,
0.14\,mag beyond their five which overlap our sample.
Conversely, again in the mean, our
Cepheids \it not \rm included in the Saha \etal\ analysis (\ie, the 12 open
circles in the lower panel of Figure \ref{fig:n3627_mod_comp}) \it
are \rm self-consistent with the overlap sample.

A clearer way in which to express the above is to simply note that the median
\it and \rm mean $\mu_\circ$ of our sample of 17 Cepheids are equivalent 
($\mu_\circ=30.04$ versus 30.06, respectively), while for the Saha \etal\
(1999) sample of 25 Cepheids, the median and mean differ by 0.19\,mag 
($\mu_\circ=30.10$ versus 30.29, respectively).  We would argue, in this case,
that the Saha \etal\ \it median \rm $\mu_\circ$ is a more robust measure of the
true DoPHOT distance to NGC~3627, in agreement with that favored by our
ALLFRAME$+$TRIAL analysis.

\subsection{NGC~3368 (M96)}
\label{n3368}

NGC~3368 was not part of the original Sandage/Saha SNe Ia program, simply
because at the time of proposing, the galaxy had not played host to a suitable
SN; this fortuitously changed on 1998 May 9, with the eruption of SN~1998bu.
While the SN itself is highly extinguished, its light-curve and photometry are
of the highest quality, and thus ideally suited for the H$_\circ$
analysis which
follows.  Tanvir \etal\ (1995), as part of their Leo~I Group study, had earlier
derived a Cepheid-based distance to NGC~3368 of $\mu_\circ=30.37\pm
0.16$ (after addition of the 0.05$\,$mag
``long-vs-short exposure'' correction - Casertano \& Mutchler 1998).

Unlike the case for the \it Key Project \rm sample,
the roll angle was not fixed from epoch-to-epoch for NGC~3368. Hence, a master
coordinate system was defined relative to the first (chronologically-speaking)
epoch Chip~2 image, and the ALLFRAME reduction performed
on all four WFPC2 chips simultaneously.

Eleven candidates were identified, their location in the resulting PL-plane
denoted by the filled circles in Figure \ref{fig:n3368}.  An
incompleteness bias, for periods less than $\sim 20$ days, not apparent in the
published data of Tanvir \etal\ (1995; Figure 3), is clearly
present.
Because Tanvir \etal\ did not include positional
information for their seven Cepheid candidates, we identified their subset,
within our larger dataset, based upon their proximity to one another in the
PL-plane.  Their sample corresponds to Cepheids C03, C05, C06, C07, C08, C10,
and C11, of Table \ref{tbl:n3368_ceph}.

\placefigure{fig:n3368}

For these seven Cepheids in common between the two studies, our ALLFRAME mean 
$V$- and $I$-band magnitudes are $0.076\pm 0.054$ and $0.122\pm 0.036$\,mag 
brighter, respectively, than those of Tanvir et~al (see Table
\ref{tbl:ceph_comp}).
The substantial differences in these mean magnitudes remains unresolved at 
present, but can be
revisited once Tanvir \etal\ publish the full photometry for their Cepheids and
local standards.  We note in passing that identifying such standards for the
ALLFRAME analysis proved challenging.  Bearing
in mind the aforementioned incompleteness bias below P=20\,d, we adopt a final
modulus based upon the 7 candidates in Table \ref{tbl:n3368_ceph} with periods
in excess of this cutoff -- \ie, $\mu_\circ=30.20\pm 0.10\,({\rm random})\pm
0.16\,({\rm systematic})$.

\subsection{NGC~5253}
\label{n5253}

NGC~5253 represents one of the two WF/PC galaxies (supplemented with some
newer WFPC2 data, in the case of NGC~5253) analyzed for this work; the
second, IC~4182 will be discussed in Section \ref{i4182}.  The reduction
of the earlier WF/PC photometry proved challenging.
The original DoPHOT
analysis led to $\mu_\circ=28.08\pm 0.2$ and E$(\vmi)=0.01\pm 0.12$ (Saha
\etal\ 1995).  The large uncertainty on the latter allows for
consistency with the measured Galactic foreground reddening 
E$(\vmi)_{\rm Gal}=0.074\pm 0.008$ (Schlegel \etal\ 1998).

NGC~5253 is unique in that it played host to two potential Type Ia SNe local
calibrators, SN~1895B and 1972E.  The quality of the photometry associated with
SN~1895B is of low quality, and therefore we do not consider it further.  The
lightcurve and photometry for SN~1972E (Hamuy \etal\ 1996) are of a decidedly
higher quality (although not of the same level as those discussed in Sections
\ref{n4639} to \ref{n3368}, primarily because observations did not begin
until 5\,d past the peak, and so one must extrapolate back to the peak using
template lightcurves)
and will be incorporated into the H$_\circ$ analysis of
Section \ref{implications}.  SN~1972E is situated well in the outskirts of
NGC~5253, $\sim 2$\,kpc from the optical center -- see Figure 1 of Saha \etal\ 
(1995). The intrinsic reddening local to SN~1972E is
measured by Suntzeff \etal\ (1999) to be
E$(\vmi)_{\rm Host}=0.013\pm 0.027$; 
if this is taken as a lower limit on the mean
intrinsic reddening, then  in combination with the
aforementioned Galactic foreground reddening, the implied total
line-of-sight reddening is E$(\vmi)_t>0.09\pm 0.03$.

Because both WF/PC and WFPC2 data exist for NGC~5253, slightly different
reduction and calibration steps were employed.  First, for the WFPC2 data, the
Stetson (1998) $I$-band calibration and CTE
corrections, based upon $\omega$~Cen and NGC~2419, were used.  Second, the
$V$-band images were not taken with the usual F555W 
filter,
but with F547M; the F547M calibration is derived from seven observations of
$\omega$~Cen taken with the GAIN=14 electronics, at odds with the GAIN=7
electronics employed in the NGC~5253 observations.  The $\omega$~Cen zeropoints
were therefore corrected with a (GAIN=7)$-$(GAIN=14) zeropoint correction
derived from F555W data on $\omega$~Cen; a separate correction was applied to
each chip.

The separately calibrated WF/PC and WFPC2 data for NGC~5253 differ by
$0.04\pm 0.01$\,mag in V, in the sense of WFPC2 being brighter, and
$0.00\pm 0.01$\,mag in I.  The weighted average was then constructed
and compared with the ground-based photometry from John Tonry (as
reported in Table 2 of Saha \etal\ 1995).  This average is heavily
weighted in favor of the WF/PC observations, simply because of the far
greater number of observations. We note also that object 3a in Saha
et~al.'s (1995) Table 2 turned out to be a clump of $\sim 6$ stars of
comparable brightness in the higher-quality
WFPC2 data, and was eliminated from the
comparison with Tonry's ground-based data. The $V$-band magnitudes
differed by $0.08\pm 0.02$, in the sense of the ground-based
photometry being brighter, and the $I$-band magnitudes differed by
$0.10\pm 0.03$ (in the same sense).  In comparison, Saha \etal\ (1995)
applied a \rm 0.14\,mag F555W zeropoint
contamination correction, and found agreement with Tonry's
ground-based photometry at the level of $0.01\pm 0.03$ ($V$-band, in
the sense of Tonry being fainter) and $0.03\pm 0.06$ ($I$-band, in
the sense of Tonry being brighter).  Because our re-analysis is anchored to the
higher-quality WFPC2 data, an \it a posteriori \rm contamination correction is
not needed.
In summary, based upon the Tonry ground-based photometry, the Saha \etal\
(1995) reference star photometry, after application of their
contamination correction, is $0.09\pm 0.04$\,mag (V) and $0.07\pm
0.07$\,mag (I) brighter than our ALLFRAME photometry, which would correspond to
an \it a priori \rm predicted ALLFRAME true modulus 0.04\,mag nearer than that
of Saha \etal\ (1995).

Seven Cepheid candidates were identified in our ALLFRAME analysis (see Table
\ref{tbl:n5253_ceph} for the relevant properties), five of which were in common
with those of Saha \etal\ (1995).
The remaining Cepheid candidates
listed in Table 4 of Saha \etal\ (1995) were examined, but were either not
recovered or (more usually) the light curve quality was not very high
or the Cepheids were located too close to the edge of the chip.
The PL fit, based upon the full ALLFRAME sample of seven Cepheids, is shown in
Figure \ref{fig:n5253}.  
The inferred apparent moduli 
-- $\mu_{\rm V}=27.95\pm 0.10$ and $\mu_{\rm I}=27.82\pm 0.08$ -- are 
significantly smaller (\ie, 0.15\,mag and 0.27\,mag in $V$ and $I$,
respectively) than those favored by Saha et~al, and at odds with the
conclusions of the previous paragraph, based upon the bright calibrator stars
in the field.
We discuss the reasons for this difference below.

\placefigure{fig:n5253}

There are five Cepheids in common between Saha \etal\ (1995) and Table
\ref{tbl:n5253_ceph} -- from our census, C02-C06 correspond to Saha et~al.'s
C4-V2, C3-V2, C4-V4
C3-V3, and C1-V2, respectively. 
Table \ref{tbl:ceph_comp} shows
that the mean magnitudes for these five Cepheids agree extremely well.
Identical moduli (both apparent and true) and
reddenings to those favored by our ALLFRAME analysis are found when using
the Saha et~al. DoPHOT data, provided the same Cepheid
sample is used.  The reason for the difference with Saha et~al.'s
results is that of the five Cepheids included in their $I$-band
analysis -- their C2-V3, C3-V6, C4-V3, C3-V2, and C4-V2 --
only the latter two were included in our final sample.

Our photometry has benefited from the WFPC2 supplement not available in
1994. 
In some cases, this has allowed us to better eliminate/include marginal Cepheid candidates from the
original Saha \etal\ analysis.  For example, C1-V2 and C3-V3 were discounted in
the $I$-band analysis by Saha \etal\ because of difficulty in assigning their 
mean magnitudes, while we found no reason to eliminate them based on their
photometry; both are clearly consistent with the PL fit of Figure
\ref{fig:n5253}, unlike the situation which faced Saha \etal\ (their Figure 5;
lower panel).  On the other hand, (i) we did not recover their C2-V3; (ii)
C4-V3 sits on a bad column; and (iii) the lightcurve of C3-V6 is clearly
non-Cepheid-like, even in their own data (their Figure 4).

We have treated the seven Cepheids in our final sample
in a manner consistent with all of the other galaxies in the \it Key 
Project\rm, and see no reason to reject any of them, based upon our 
ALLFRAME$+$TRIAL analysis.

We note also that 
our inferred mean reddening (E$(\vmi)=0.14\pm
0.05$), is larger than that
found by Saha \etal\ (1995) (E$(\vmi)=0.01\pm 0.12$).  
Our higher value of reddening is in better agreement with other
estimates of the reddening (recall,
the Galactic foreground reddening is E$(\vmi)\approx 0.08$ --
Schlegel \etal\ 1998). In addition, over the inner $25^{\prime\prime}\times
25^{\prime\prime}$ (half of our Cepheids lie within $20^{\prime\prime}$ of this
region), the total line-of-sight reddening (from 
$H\alpha/$H$\beta$ -- Calzetti \etal\ 1997) is E$(\vmi)\approx 0.22$.

In summary, 
our adopted distance modulus for NGC~5253, incorporating the full error budget,
is $\mu_\circ=27.61\pm 0.11\,({\rm random})\pm 0.16\,({\rm systematic})$.
The main difference  from that of the published Saha \etal\ (1995)
analysis (\ie, $\mu_\circ=28.08\pm 0.2$)
is due to the difference in adopted Cepheid samples. The small scatter
in our ALLFRAME PL relation and individual light curves, combined with the 
larger inferred reddening value, plus larger time baseline of the data
analyzed here (in a manner consistent with all previous \it Key
Project \rm galaxies), all give us confidence in the Cepheid sample
adopted in this paper. 

An independent check of our (and
Saha et~al.'s) conclusions is provided by the Sakai (1999) distance
determination based upon the magnitude of the tip of the red giant branch
(TRGB); a
preliminary analysis, using only NGC~5253 halo stars, places the tip at
$I$=23.99.  For a tip absolute magnitude M$_{\rm I}\equiv -4.0$
(Ferrarese \etal\ 1999), and a Galactic foreground extinction
A$_I=0.11$\,mag (Schlegel \etal\ 1998),
the inferred TRGB distance to NGC~5253
would be $\mu_\circ^{\rm T-RGB}=27.88$.  This value for $\mu_\circ^{\rm T-RGB}$
lies between Saha \etal\ ($\mu_\circ=28.08$)
and our current result ($\mu_\circ=27.61$), in (dis)agreement with
both of the Cepheid-based distance determinations at the $1-1.5\sigma$ 
level.\footnote{As rightly pointed out by the referee, one can use the SNe
themselves to estimate the distance to NGC~5253.  Using SN~1937C,
1981B, 1990N, and 1998bu as the calibrators, and following the methodology to
be described later in Section \ref{hamuy}, a distance modulus of
$\mu_\circ=27.83$ can be derived for SN~1972E which, like the Sakai (1999)
T-RGB distance to the host galaxy, lies half-way between our new Cepheid
distance ($\mu_\circ=27.61$) and the original Saha \etal\ (1995) result
($\mu_\circ=28.08$).}

An ultimate resolution to the discrepancy between each of these approaches to
the distance of NGC~5253 is still forthcoming.  We should stress
that the larger random uncertainty associated with our ALLFRAME$+$TRIAL 
distance ensures that under the weighting scheme employed in 
Section~\ref{implications}, NGC~5253, by itself, does not overly influence 
our final result.

In
passing, we note that the formal Saha \etal\ solution of $\mu_\circ=28.08$ was
not adopted in the subsequent Sandage \etal\ (1996) and Saha \etal\ (1997)
H$_\circ$ analyses; instead, a true modulus of $\mu_\circ=28.00$ appears
to have been adopted. 

\subsection{IC~4182}
\label{i4182}

IC~4182 is the host galaxy for the SNe~Ia calibrator SN~1937C.  While the
photographic calibration of the SN~1937C lightcurve is of lower
quality (in comparison with the SNe of Sections \ref{n4639} to \ref{n3368}),
the quality of the lightcurve itself is high, and for that reason is included
in the H$_\circ$ analysis of Section \ref{implications}.

The original DoPHOT reduction for this galaxy
uncovered 27 high-quality Cepheids (Saha \etal\ 1994); 
the published moduli were
$\mu_V=28.31\pm 0.05$ and $\mu_I=28.42\pm 0.05$.  The inferred
reddening E(\vmi)=$-0.11\pm 0.07$ is clearly unphysical, which led Saha \etal\ 
to assume $\mu_\circ\equiv(\mu_V+\mu_I)/2=28.36\pm 0.09$.

For the ALLFRAME analysis, the calibration adopted parallels that of Section
\ref{n5253} and Stetson (1998), although in this case there were no 
supplementary WFPC2
data.  Twenty-eight Cepheid candidates were identified (properties as noted in
Table \ref{tbl:i4182_ceph}), 27 of which were in common with those in Table 3
of Saha \etal\ (1994). As can be seen from inspection of Table
\ref{tbl:ceph_comp}, there is a difference in the photometry for these
Cepheids in common -- the Stetson (1998)
ALLFRAME mean magnitudes are 0.038\,mag ($V$)
and 0.119\,mag ($I$)
brighter than the Saha \etal\ DoPHOT magnitudes.
The PL relations, and associated fits, for the 28 Cepheids of Table
\ref{tbl:i4182_ceph} are shown in Figure \ref{fig:i4182}.  

\placefigure{fig:i4182}

Our adopted ALLFRAME$+$TRIAL true modulus, with the full error accounting,
of $\mu_\circ=28.36\pm 0.08\,({\rm random})\pm 0.16\,({\rm systematic})$ (Table
\ref{tbl:distances}) is identical to the Saha \etal\ (1994) result of
$\mu_\circ=28.36\pm 0.09$, although this excellent agreement in true modulus is
somewhat misleading.  Specifically, our re-analysis led to a formal value of
the reddening of E(\vmi)=$-0.03\pm 0.03$, while Saha \etal\ found
E(\vmi)=$-0.11\pm 0.04$.  To avoid any bias, our quoted true modulus includes
the correction for the small, but negative, reddening implied by our Cepheid
data; Saha \etal\ applied no such correction.  We note in passing that the
Galactic foreground reddening along the line-of-sight
to IC~4182 is E(\vmi)=$0.01$ (Schlegel \etal\ 1998), consistent with
our Cepheid-derived mean reddening. 

Preliminary reduction of ground-based photometry of IC~4182, provided by 
M.~Pierce, confirms the validity of our adopted zeropoints at the level of 
$\simlt 0.1$\,mag, within the accuracy of the ground-based data.  
Further confirmation of the ground-based tie-in is underway.  

\subsection{NGC~4496A}
\label{n4496a}

In the case of SN~1960F, the light-curve shape and photographic
calibration are of low quality (see Schaefer 1996).  Indeed, Suntzeff \etal\
(1999) and earlier papers in their series, do not include this SN in
their H$_\circ$ analysis because of its questionable quality. For
completeness, however, we have re-analyzed the data for its host
galaxy, NGC~4496A, and provide solutions for H$_\circ$ with and
without its inclusion in the Hubble diagram zero points, in Section
\ref{implications}.

Based upon 44 Cepheids with P$>$18\,d, color $0.6<(\vmi)<1.1$, and quality
index Q$\ge 4$, Saha \etal\ (1996b) found 
$\mu_\circ=31.03\pm 0.14$ and E(\vmi)=0.04$\pm$0.06.  Our ALLFRAME$+$TRIAL
re-analysis uncovered 94 Cepheids (Table \ref{tbl:n4496a_ceph}), the PL
relations for which are shown in Figure \ref{fig:n4496a}.

\placefigure{fig:n4496a}

Restricting ourselves to the 51 Cepheids with
P$>$25\,d, to minimize the effects of incompleteness
bias (apparent in Figure \ref{fig:n4496a}), our distance modulus for
NGC~4496A is
$\mu_\circ=31.02\pm 0.07\,({\rm
random})\pm 0.16\,({\rm systematic})$.
Our modulus is consistent with the published Saha \etal\ (1996b)
value.

In their H$_\circ$
analyses, Sandage \etal\ (1996) and Saha \etal\ (1997) do not adopt
their published modulus (\ie, $\mu_\circ=31.03\pm 0.14$), and instead adopt the
apparent $V$-band modulus (\ie, $31.13\pm 0.10$) in its place.  
For consistency with standard practice in the \it Key Project\rm, we employ
the true modulus $\mu_\circ$, and not the apparent modulus $\mu_V$, in
Section \ref{implications}.

\subsection{Summary}
\label{Summary_Distances}

Table \ref{tbl:distances} summarizes the analysis and discussion of Sections
\ref{n4639}-\ref{n4496a}.  Column (2) lists the new ALLFRAME$+$TRIAL distances
derived in the current study, while those values originally published in the
Tanvir \etal\ (1995) and Saha \etal\ (1994,1995,1996ab,1997,1999) papers are
provided in column (4).  The unweighted mean offset between the two datasets is
$<\mu_\circ({\rm ALL})-\mu_\circ({\rm pub})>_{\rm u}=-0.16$\,mag, while the 
weighted mean is $<\mu_\circ({\rm ALL})-\mu_\circ({\rm pub})>_{\rm w}=-0.12\pm
0.07$\,mag -- \ie, a discrepancy at the $\sim$2$\sigma$ level.  Neglecting
IC~4182, since the Saha \etal\ (1994) uncertainty of $\pm 0.09$\,mag appears to
be too small,\footnote{Despite
the inherent difficulties in calibrating the older WF/PC
data, and the apparent $I$-band problems which led Saha \etal\ (1994) to infer a
highly negative reddening for the IC~4182 Cepheids, their quoted true modulus
uncertainty ($\pm 0.09$\,mag) is significantly smaller than that associated with
any of their other six galaxies (see column (4) of Table \ref{tbl:distances}),
and is also smaller than the uncertainty associated with the true modulus of
the LMC, upon which the relative distance is based.} yields
$<\mu_\circ({\rm ALL})-\mu_\circ({\rm pub})>_{\rm u}=-0.19$\,mag and
$<\mu_\circ({\rm ALL})-\mu_\circ({\rm pub})>_{\rm w}=-0.17\pm 0.08$\,mag.
Further, neglecting NGC~5253, due to its extreme outlier 
status, results in
$<\mu_\circ({\rm ALL})-\mu_\circ({\rm pub})>_{\rm u}=-0.13$\,mag and
$<\mu_\circ({\rm ALL})-\mu_\circ({\rm pub})>_{\rm w}=-0.13\pm 0.08$\,mag, 
similar to what was found using the full sample of seven galaxies.
Using the result based upon this full sample, $\sim$1/2 of the 0.12\,mag
discrepancy can be traced
to unresolved photometric calibration issues (see Table \ref{tbl:ceph_comp}),
with the remainder ascribed to
the different Cepheid samples and PL-fitting methodology employed.
Again, as there exist serious concerns
regarding either the DoPHOT Cepheid sample or photometry, for each of the seven
galaxies analyzed, applying an ad hoc \it a posteriori \rm
``averaging'' of the DoPHOT and ALLFRAME distance moduli is \it not \rm
recommended.  Consistent with the general \it Key Project \rm practice of
adopting the ALLFRAME results, we do so here as well.

\placetable{tbl:distances}

\placetable{tbl:supernovae}

Column (3) of Table \ref{tbl:distances} presents the
true moduli favored by the standard \it Key Project \rm
PL-fitting methodology employed in the current study, but using the tabulated
Saha \etal\ and Tanvir \etal\ photometry, while column (5) shows the true
moduli used by Sandage \etal\ (1996) and Saha \etal\ (1997) in their \ho\
analyses (c.f. column 4).  Table \ref{tbl:supernovae} lists the adopted
photometry and SNe light curve characteristics for each of these nearby
calibrators; the peak luminosities of the SNe (columns 7-9) were calculated 
using the $\mu_\circ({\rm ALL})$ values of Table \ref{tbl:distances}.
How the adoption of these subsets of true moduli (Table \ref{tbl:distances})
and SNe photometry (Table \ref{tbl:supernovae})
impacts upon the derivation of H$_\circ$ 
is explored in Section \ref{implications}.

\section{The Implications for the Hubble Constant}
\label{implications}

Given the distance moduli derived from our ALLFRAME$+$TRIAL
re-analysis of the
local Type Ia SNe-host galaxies of Table \ref{tbl:distances} (column 2), we are
now in a position to calibrate the distant sample of
SNe comprising the C\'alan-Tololo-based Hubble diagrams.  Two 
approaches are considered here -- one based upon the adoption of an absolute
magnitude-peak luminosity correlation correction (Section \ref{hamuy},
paralleling Suntzeff \etal\ 1999), and one
with no correction (Section \ref{sandage}, paralleling 
Saha \etal\ 1997).  This is not meant
to be an exhaustive replication of all previous Type Ia SNe-based
H$_\circ$ determinations, nor is one warranted;  for convenience, we have
chosen the two extremes in the treatment of SNe lightcurves.  
While a variety of different approaches have been taken
(\eg, Phillips 1993; Tammann \& Sandage 1995;
Hamuy \etal\ 1996; Saha \etal\ 1997; Riess \etal\ 1998; Phillips \etal\ 
1999; Jha \etal\ 1999), 
here we are simply interested in exploring the effects of our new 
distance
determinations upon the zero points of the two extrema.
Because all
empirically-based studies listed in Table \ref{tbl:H0_Ia} rest squarely upon
the zero point defined by the same local calibrators, 
the differential effect is essentially identical, regardless of
the details in the treatment of the distant sample.

\subsection{Suntzeff \etal\ (1999)}
\label{hamuy}

We employ the reddening-corrected 
M$_{\it BVI}^{\rm max}-\Delta m_{15}(B)$ Hubble relations described by
Suntzeff \etal\ (1999) and Phillips \etal\ (1999).  Following Suntzeff et~al., 
we adopt the zero points, coefficient uncertainties, and Hubble diagram fit
dispersions derived from the full sample of 35 C\'alan-Tololo/CfA Type Ia SNe;
coefficient uncertainties and dispersions were kindly provided by Mark 
Phillips. The fits to the decline rate versus peak luminosity 
relation from Phillips \etal\ (1999) were derived from the 
low-extinction (\ie, E(\bmv)$<$0.05) subset of 18
Type Ia SNe in the C\'alan-Tololo/CfA sample.  While the Suntzeff \etal\ (1999)
fits included corrections
for both Galactic and host galaxy reddening, the Phillips
\etal\ fits include only foreground Galactic reddening, although it should be
stressed that
the zero points for these two sets agree to $<1\sigma$.
Our adopted relations are represented by the following expressions
for H$_\circ$ in $B$, $V$, and $I$.
\begin{eqnarray}
\log{\rm H}_\circ(B) & = & 0.2\big\{{\rm
M^{max}_B}-0.720(\pm0.459)\big[\Delta m_{15}(B)_t-1.1\big]- \nonumber \\
&& 1.010(\pm0.934)\big[\Delta m_{15}(B)_t-1.1\big]^2
+ 28.653(\pm0.042)\big\}, \label{eq:Hamuy_B} \\
\log{\rm H}_\circ(V) & = & 0.2\big\{{\rm
M^{max}_V}-0.672(\pm0.396)\big[\Delta m_{15}(B)_t-1.1\big]- \nonumber \\
&& 0.633(\pm0.742)\big[\Delta m_{15}(B)_t-1.1\big]^2
+ 28.590(\pm0.037)\big\}, \label{eq:Hamuy_V} \\
\log{\rm H}_\circ(I) & = & 0.2\big\{{\rm
M^{max}_I}-0.853(\pm0.214)\big[\Delta m_{15}(B)_t-1.1\big]+28.219(\pm0.034)\big\}, \label{eq:Hamuy_I}
\end{eqnarray}
\noindent
where
\begin{equation}
\Delta m_{15}(B)_t = \Delta m_{15}(B)_{\rm obs} + 0.1\,
{\rm E}(\bmv)_t.  \label{eq:red}
\end{equation}
\noindent
E$(\bmv)_{t}$ represents the sum of E$(\bmv)_{\rm Gal}$ (\ie, foreground
Galactic reddening) and E$(\bmv)_{\rm Host}$ (\ie, intrinsic SN reddening) from
Table 5 of Suntzeff \etal\ (1999).  For SN~1989B, we assign 
E$(\bmv)_{t}=0.37\pm
0.03$ (Wells \etal\ 1994), for SN~1960F, E$(\bmv)_{t}=0.01\pm0.01$ (Schaefer
1996), and for SN~1974G, E$(\bmv)_{t}=0.16\pm 0.07$ (Schaefer 1998).  The
dispersions for the fits of equations \ref{eq:Hamuy_B}-\ref{eq:Hamuy_I} are
$\sigma_B=0.16$, $\sigma_V=0.16$, and $\sigma_I=0.17$
(Phillips 1999, priv comm).

Suntzeff \etal\ (1999) provide a zeropoint to equations 
\ref{eq:Hamuy_B}-\ref{eq:Hamuy_I} using the following
absolute distances to the host
galaxies of five Type Ia SNe -- (1) $\mu_\circ=32.03\pm 0.22$ for SN~1990N in
NGC~4639; (2) $\mu_\circ=31.10\pm 0.13$ for SN~1981B in NGC~4536; (3)
$\mu_\circ=30.37\pm 0.16$ for SN~1998bu in NGC~3368; (4) $\mu_\circ=28.31\pm
0.11$ for SN~1937C in IC~4182; and (5) $\mu_\circ=27.92\pm 0.08$ for SN~1972E
in NGC~5253.  The adopted moduli were then paired with the corresponding SN's
apparent peak magnitude (columns 3-5 of Table 4) and intrinsic reddening (sum
of columns 10 and 11 of Table 5 in Suntzeff \etal\ 1999), in order to derive
M$_{\it BVI}^{\rm max}$.  These peak luminosities and associated decline rates
$\Delta m_{15}(B)$ were then used in conjunction with equations
\ref{eq:Hamuy_B}-\ref{eq:Hamuy_I} to derive a color- and SN-dependent
H$_\circ$.  The weighted average of the resulting set of Hubble constants was
H$_\circ=64.0\pm 2.2$\,km\,s$^{-1}$\,Mpc$^{-1}$.

Before demonstrating how our new distances to the galaxies in question (column
2 of Table \ref{tbl:distances}) impact upon the Suntzeff \etal\ (1999) value of
H$_\circ=64.0\pm 2.2$ \hounits, we note the following:

First, Suntzeff \etal\ use a ratio of total-to-selective absorption
of R$_V=3.1$, while all distances quoted by the \it Key Project\rm, including
those of Table \ref{tbl:distances}, assume R$_V=3.3$.  In the analysis which
follows, whenever the Suntzeff \etal\ (1999) Hubble relations (\ie, equations 
\ref{eq:Hamuy_B}-\ref{eq:red}) are employed (Sections \ref{hamuy}, 
\ref{metallicity}, 
and \ref{flow}), we adjust the true moduli (and associated Cepheid reddenings)
of Table \ref{tbl:distances}, to ensure consistency with Suntzeff et~al.'s use
of R$_V=3.1$.  We stress that the effect upon H$_\circ$ is $<<$1\%, and
entirely accounted for in our error budget (entry R2 of Table
\ref{tbl:error}).
Second, Suntzeff \etal\ correctly adopt $\mu_\circ=30.37\pm 0.16$ for NGC~3368,
and not the published $\mu_\circ=30.32\pm 0.16$ (Tanvir \etal\ 1995); the
additional 0.05\,mag corresponds to the ``long-versus short'' effect, not
included by Tanvir et~al.
Third, Suntzeff \etal\ use (slightly) modified absolute moduli for IC~4182 and
NGC~5253, from those published by Saha \etal\ (1994) and Saha \etal\ (1995),
respectively.  While the published values were $\mu_\circ=28.36\pm 0.09$
(IC~4182) and $\mu_\circ=28.08\pm 0.2$ (NGC~5253), Suntzeff \etal\ used
$\mu_\circ=28.31\pm 0.11$ (IC~4182) and $\mu_\circ=27.92\pm 0.08$ (NGC~5253).
In the case of NGC~5253, a long-versus-short correction of
0.05\,mag was applied
to an apparent $V$-band modulus of $\mu_V=28.08\pm 0.07$ , followed by the subtraction of a total
line-of-sight extinction of A$_V=0.21$\,mag, resulting in a $\mu_\circ$
which is 0.16\,mag nearer than that favored by Saha \etal\ (1995).
This
differs (mildly) from the Saha \etal\ (1995) published $V$-band modulus of
$\mu_V=28.10\pm 0.07$.
Since there is no extant evidence for a
long-versus-short effect in the older WF/PC data (\ie, for our
purposes, IC~4182 and NGC~5253),  such a correction is not necessary.

Our re-analysis of the Saha \etal\ and Tanvir \etal\ galaxies allows us to
circumvent some of the problems encountered by
Suntzeff \etal\ (in particular, the
assumptions that had to be made in order to derive a suitable $\mu_\circ$ for
IC~4182 and NGC~5253).  Applying the self-consistent distances listed in column
2 of Table \ref{tbl:distances}, in lieu of those at the disposal of Suntzeff
et~al., along with
the SNe photometry of Table \ref{tbl:supernovae}, to the Hubble relations of
equations \ref{eq:Hamuy_B}-\ref{eq:Hamuy_I}, leads to the set of SN- and
color-dependent Hubble constants shown in Table \ref{tbl:hamuy}.  A sample
error budget for SN~1981B in NGC~4536, is provided in Table \ref{tbl:error},
similar to that of Table~4 of Hamuy \etal\ (1996).

\placetable{tbl:hamuy}

\placetable{tbl:error}

Using the full sample of eight calibrators in Table \ref{tbl:hamuy} leads to
a weighted, color-independent, Hubble constant of H$_\circ=67.7\pm 
1.7\,({\rm random}) \pm 5.0\,({\rm systematic})$
\hounits.  Because of the previously documented shortcomings of SN~1960F
and 1974G, we favor a Hubble constant tied to the remaining six
calibrators (\ie, the five employed by Suntzeff \etal\ 1999, supplemented
by SN~1989B in NGC~3627), although we should stress that numerically the
results are identical.  Therefore,
our favored result is
\begin{equation}
{\rm H}_\circ = 67.9\pm 1.9\,({\rm random})\pm 5.0\,({\rm systematic}) 
\qquad{\rm km\,s^{-1}\,Mpc^{-1}.} 
\label{eq:ho_bkg}
\end{equation}
\noindent
Comparing equation \ref{eq:ho_bkg}
with the Suntzeff \etal\ 
(1999) result of H$_\circ=64.0\pm 2.2$ \hounits\
demonstrates that our new ALLFRAME$+$TRIAL-based distances
result in a 6.1\% increase in the predicted Hubble constant.

\subsection{Saha \etal\ (1997)}
\label{sandage}

In contrast with the Suntzeff \etal\ (1999) methodology of Section \ref{hamuy},
Saha \etal\ (1997), Sandage \etal\ (1996)  do not apply a 
peak luminosity-decline rate correction aimed at reducing the scatter of the
Type Ia SNe Hubble diagram.  Their adopted fits, which are based upon a 
combination of the \ct and Asiago (Patat 1995) surveys, are
\begin{eqnarray}
\log{\rm H}_\circ(B) & = & 0.2\big\{{\rm
M^{max}_B}+28.265(\pm0.045)\big\} \label{eq:saha_B} \\
\log{\rm H}_\circ(V) & = & 0.2\big\{{\rm
M^{max}_V}+28.295(\pm0.040)\big\} \label{eq:saha_V}
\end{eqnarray}
\noindent
and are the direct analog of equations 
\ref{eq:Hamuy_B} and \ref{eq:Hamuy_V}.

If we now take the peak
absolute magnitudes for the six 
calibrators used in Section \ref{hamuy}, 
from Table 6 of Saha et~al., we find the weighted means to be
M$_B^{\rm max}=-19.48\pm 0.08$ and M$_V^{\rm max}=-19.48\pm 
0.07$. While Saha \etal\ (1997) use SN~1895B
in their H$_\circ$ analysis, we have not, due to 
the larger errors in its photographic photometry.
The good agreement between these mean magnitudes 
and the ones we derived in Section \ref{hamuy} (see the final two entries 
of Table \ref{tbl:supernovae}), despite the $0.12\pm 0.07$\,mag systematic
offset in the absolute distances (recall Table \ref{tbl:distances})
is largely a fortuitous one.  It comes about due to the fact that Saha
\etal\ (1997) assume the observed SNe peak magnitudes reflect the unextinguished
values (except in the cases of SN~1981B and SN~1989B), whereas Suntzeff \etal\ 
(1999) solve for both the intrinsic and Galactic foreground reddening.
Coupled with equations \ref{eq:saha_B} and
\ref{eq:saha_V}, these lead to H$_\circ(B)=57.2\pm 2.2$\,
km\,s$^{-1}$\,Mpc$^{-1}$ and H$_\circ(V)=57.9\pm 2.0$\,
km\,s$^{-1}$\,Mpc$^{-1}$, respectively.  The weighted mean of these $B$- and
$V$-band Hubble constants is H$_\circ = 57.6\pm 1.5$ \hounits.
\noindent
Saha \etal\ (1997) use slightly different distances to the majority of their
calibrators, in comparison with those published in their individual galaxy
papers; adopting these
published distances to the galaxies in question leads
to H$_\circ=58.3\pm 1.7$ \hounits, a 1.2\% increase over that above.
In comparison, 
adopting our best estimate of the distances to the six Sandage/Saha
calibrators, based upon their
original published DoPHOT photometry (column 3 of Table \ref{tbl:distances}),
would yield H$_\circ=61.5\pm 1.4\,({\rm random})\pm 4.5\,({\rm systematic})$
\hounits\, (entry 7 of Table \ref{tbl:H0_bkg}).

If we were to now replicate the analysis of Saha \etal\ (1997), again
using their SNe photometry and reddening parameters (Table 6 of Saha
et~al.), but now adopting the new ALLFRAME$+$TRIAL
distance moduli of Table \ref{tbl:distances} (column 2),
the predicted mean peak magnitudes would be M$_B^{\rm
max}=$M$_V^{\rm max}=-19.27\pm 0.06$  \ie,  the
mean $B$- and $V$-band peak magnitudes both decrease by 
$\sim$0.21\ mag.  

In conjunction with
equations \ref{eq:saha_B} and \ref{eq:saha_V}, these revised mean magnitudes
lead to Hubble constants of H$_\circ(B)$=$62.1\pm 2.0\,({\rm random})\pm
4.6\,({\rm systematic})$
\hounits and H$_\circ(V)$=$63.2\pm 1.9\,({\rm random})\pm 4.6\,({\rm 
systematic})$ \hounits.  
The weighted mean of these values is H$_\circ=62.7\pm 1.4\,({\rm random})\pm
4.6\,({\rm systematic})$ \hounits\, (entry 5 of Table \ref{tbl:H0_bkg}).  This 
value is directly comparable to the H$_\circ=57.6\pm 1.5$ \hounits\ alluded to
earlier, in that the adopted Hubble diagram (equations
\ref{eq:saha_B} and \ref{eq:saha_V}) and SN lightcurve photometry (Table
6 of Saha \etal\ 1997) were chosen to be identical -- the only difference here 
is the adopted absolute distances to the host galaxies in question.   The
adoption of the new distance moduli leads to an upward revision of the Saha
et~al. (1997) H$_\circ$ of 8.9\%.

Saha \etal\ (1997) have noted that when correcting for the decline
rate adopted by Suntzeff \etal\ (1999), the calibrators become
$\sim$0.1\ mag more luminous than the distant sample. Suntzeff \etal\
(1999, and earlier papers) suggest that such an anomaly may be real,
reflecting the morphological type (and stellar population) mismatch
between nearby and distant host galaxies.  On balance, as summarized
earlier, the case for a decline rate correction seems strong and hence
our adoption of such an approach.

Saha \etal\ use a different Type Ia SNe sample for deriving their
Hubble Diagram.  Their derived decline rate-peak luminosity relationships (in
both $B$ and $V$) are $\sim$10\% shallower than those derived by Suntzeff
\etal\ (1999)
and Phillips \etal\ (1999).  Using these Saha \etal\ fits (their
equations 23 and 24, with the associated zero points) with the new
ALLFRAME$+$TRIAL distances to the same six calibrators employed above, 
leads to H$_\circ$=$65.5\pm 1.5\,({\rm random})\pm 4.8\,({\rm systematic})$ 
\hounits, which is 4.5\% greater than the
value derived in the absence of any M$_{\it BV}^{\rm max}-\Delta m_{\rm
15}(B)$ correction (\ie, entries 5 and 6 of Table \ref{tbl:H0_bkg}).  
In comparison, the
steeper slope favored by Suntzeff \etal\ (1999) results in an 8.3\%
increase in H$_\circ$ (\ie, entries 1 and 5 of Table \ref{tbl:H0_bkg}).

\subsection{The Effect of Metallicity in Determining H$_\circ$}
\label{metallicity}

The results of Sections \ref{hamuy} and \ref{sandage} were 
assuming no metallicity dependence of the Cepheid PL
relation.  The dependence of the Cepheid PL relation on metallicity
has not yet been well--established (\eg, Musella 1999; Storm,
Carney \& Fry 1999; Nevalainen \& Roos 1998; Saio \& Gautschy 1998;
Webb \etal\ 1998; Kennicutt \etal\ 1998; Sasselov \etal\ 1997;
Gould 1994; Madore \& Freedman 1990), with perhaps the best empirical
determination of the magnitude of such an effect to date being that due to 
Kennicutt \etal\ (1998).  

Kennicutt \etal\ performed a differential analysis of the PL
relations in two fields of M101, spanning a range of $\sim 0.7$\,dex in oxygen
abundance, and found
$\Delta\mu_\circ/\Delta[{\rm O/H}]=-0.24\pm 0.16$\,mag\,dex$^{-1}$.  In other
words, for any galaxy whose metallicity exceeds that of the reference galaxy
- \ie, the LMC, with $12+\log({\rm O/H})=8.5$ -- by an order of magnitude, 
simply
determining the absolute distance modulus via offsetting the apparent PL
relations from those of the LMC, could lead to $\mu_\circ$
being underestimated by 0.24\,mag.

For each of the six high- and intermediate-quality calibrating SNe of Table
\ref{tbl:galaxies}
Kennicutt \etal\ (1999) have now shown that, in 
the mean, the metallicity of the Cepheid fields, as measured by \ion{H}{2} 
abundances, is 0.31\,dex greater than that of the LMC.  Recalling the Kennicutt
\etal\ (1998) result of $\Delta\mu_\circ/\Delta[{\rm O/H}]=-0.24\pm
0.16$\,mag\,dex$^{-1}$, this 
$+$0.31\,dex metallicity offset from the LMC implies
a typical correction factor of $\Delta\mu_\circ=+$0.07\,mag 
needs to be applied to the ensemble of calibrators.  

Using the \ion{H}{2} abundances from Table \ref{tbl:galaxies}, we applied the
appropriate Kennicutt \etal\ (1998) correction to each of the individual
calibrators, and then followed the identical decline rate-peak luminosity
analysis described in Section \ref{hamuy}.
This leads to a predicted Hubble constant of
H$_\circ = 64.9\pm 1.8\,({\rm random})\pm 4.9\,({\rm systematic})$ \hounits.
\noindent 
In other words, 
applying a correction to the
absolute distance moduli, based on Kennicutt \etal\ (1998) relationship,
results in a reduction in our adopted H$_\circ$ of 4.4\% (entries 1 and 3
of Table \ref{tbl:H0_bkg}).  These two
results agree to within $\sim$1$\sigma$.

\subsection{The Effect of the Local Flow Field in Determining H$_\circ$}
\label{flow}

The Type Ia supernova sample is sufficiently distant that use of the Cosmic Microwave Background
(CMB) reference frame is the standard to which the observed Type Ia SNe
host galaxy heliocentric velocities are transformed.
Here we test the
assumption that correcting velocities for the local flow field
does not introduce any significant systematic offset in the derived H$_\circ$.
The CMB frame  should, of course, be appropriate, as all but 5 of the 29
\ct SNe have recessional velocities in excess of 6000\,km\,s$^{-1}$ (\ie, a
regime where peculiar velocities are $\simlt 5\rightarrow 10$\% the recessional
velocity); for the nearby SNe, though, the peculiar velocities can be
significant, so an effort has been made to estimate them.

To that end, 
we have employed the
three-attractor (Virgo$+$Great Attractor$+$Shapley) model
described by Mould \etal\ (1999b), in order to correct the observed
heliocentric velocities for the resultant flow field solution.  
We assume Local Group infall velocities
of 200\,km\,s$^{-1}$, 400\,km\,s$^{-1}$, and 85\,km\,s$^{-1}$, for Virgo, the
Great Attractor, and Shapley, respectively.
Within this framework,
the adopted infall velocities  reproduce the
amplitude of the CMB dipole ($\sim$630\,km\,s$^{-1}$), but with only these
three attractors, the direction of Local Group motion is $\sim$20$^\circ$ away
from the CMB direction.  As we are simply interested in assessing the magnitude
of the effect, at this stage, fine-tuning the model to improve this vector
offset was deemed unnecessary.

Starting from 29 \ct SNe (Hamuy \etal\ 1996; Table 1), we adopted the Hamuy
\etal\ (1996) color cut to eliminate the three reddest SNe (SN~1990Y, 1992K,
and 1993H), and also eliminated the three
which are either missing (SN~1990T and 1991S) from \tt zcat \rm (Huchra 1999),
or have \tt zcat \rm velocities which are discrepant with those tabulated by
Hamuy et~al (SN~1993ag).  To this final sample of 23 Type Ia SNe, we performed
a weighted least squares fit of the form
\begin{equation}
{\rm m}_{\rm max}-5\log v = c_1\big[\Delta m_{15}(B)-1.1\big] + c_2
\label{eq:fit}
\end{equation}
\noindent
including errors in m$_{\rm max}$ and $\Delta m_{15}(B)$, but neglecting
those in recessional velocity $v$.
Restricting ourselves to the peak 
$V$-band magnitude $V_{\rm max}$ and the CMB reference frame (\ie, $v\Rightarrow
v_{\rm CMB}$ in equation \ref{eq:fit}), for the adopted sample of 23 SNe, 
we find
$c_1=0.930\pm 0.096$ and $c_2=-3.376\pm 0.029$, with a dispersion $\sigma_V=
0.17$.  Following the error analysis outlined in Table \ref{tbl:error}, and
using the same six calibrators employed in Section \ref{hamuy}, this yields a
$V$-band Hubble constant of
${\rm H}_\circ(V) = 62.0\pm 2.8\,({\rm random})\pm 4.6\,({\rm systematic})$
\hounits.
Conversely, if we now take the output of our three-attractor flow field model
and replace $v_{\rm CMB}$, with the derived flow field velocities $v_{\rm FF}$,
the coefficients for equation \ref{eq:fit} become 
$c_1=1.061\pm 0.108$ and $c_2=-3.413\pm 0.031$, with a dispersion $\sigma_V=
0.20$.  The resultant $V$-band Hubble constant is then
${\rm H}_\circ(V) = 63.3\pm 3.1\,({\rm random})\pm 4.7\,({\rm systematic})$
\hounits.

Comparison of the above results demonstrates
that H$_\circ(V)$ increases by only 2.1\% in going from the CMB to
flow field reference frame.  The same conclusion holds for H$_\circ(B)$
and H$_\circ(I)$, not surprisingly.  It is re-assuring to confirm that
the two reference frames lead to Hubble constants which are consistent at the
$<1\sigma$ level, as we had assumed, \it a priori\rm.

\subsection{A Comment Regarding Type Ia Supernovae and the Fornax Cluster}
\label{fornax}

Under
the assumption that the Fornax distance was defined by the Silbermann
\etal\ (1999) Cepheid distance to NGC~1365 - \ie, 
$\mu_\circ=31.31\pm 0.20\,({\rm
random})\pm 0.18\,({\rm systematic})$ - the suggestion had been made (\eg,
Figure 4 of Suntzeff \etal\ 1999) that the (two) Fornax Cluster
Type Ia SNe appeared to
lie $\sim$0.35\,mag behind NGC~1365 (a discrepancy, albeit only at the
$\sim$1.5$\sigma$ level).  
Indeed, the same apparent discrepancy with this earlier Cepheid distance had
been noticed in surface brightness fluctuation
(SBF, 
Ferrarese \etal\ 1999) and Fundamental Plane (FP, Kelson \etal\ 1999) Fornax
studies.
With the addition of new Cepheid distances to
NGC~1326A (Prosser \etal\ 1999) and 1425 (Mould \etal\ 1999a), 
though, the \it Key Project\rm's favored Fornax distance
is now $\mu_\circ=31.59\pm 0.04\,({\rm random})$ (Ferrarese \etal\ 1999), 
reducing the Cepheid-SNe distance discrepancy from the earlier
$\sim$1.5$\sigma$, to $<$1$\sigma$.  Further, direct, Fornax distance
determinations, including Sandage/Saha's
upcoming attempt at uncovering Cepheids in NGC~1316,
will be an important aid in determining whether the earlier Cepheid versus
SBF/FP distance ``dichotomy'' has truly been resolved.  Regardless, as far as a
Type Ia SNe-based determination of H$_\circ$ is concerned, the inclusion of
these two additional Fornax SNe has little effect upon the derived result, save
for the expected reduction in the final random uncertainty (Madore \etal\
1999).

\subsection{Summary}
\label{summ_H0}

Table \ref{tbl:H0_bkg} summarizes the results of Sections
\ref{hamuy}-\ref{flow}, illustrating the dependence of the derived Hubble
Constant to the important constituents of the analysis, including the adopted
peak luminosity-light curve shape correction, number of local calibrators,
Cepheid PL-metallicity, and reference frame.  

\placetable{tbl:H0_bkg}

It is important to stress that  all  of the scenarios based upon our
uniform ALLFRAME$+$TRIAL analysis (all but the final entry of Table
\ref{tbl:H0_bkg}) are consistent, at the $<1.5\sigma$ level, with our final,
adopted, result of H$_\circ=68\pm 2\,({\rm random})\pm 5\,({\rm
systematic})$ \hounits (\ie, the initial entry of Table \ref{tbl:H0_bkg}).
It is reassuring that this result is so robust,
considering, for example,
the different approaches adopted in the treatment (or lack thereof)
of the SNe light curve shapes (contrast the first and fifth entries of Table
\ref{tbl:H0_bkg}).

\section{Conclusions}
\label{summary}

We have re-derived distances to seven
Type Ia SNe-host galaxies, using archival
\it HST \rm data, collected by
two different, non-\it Key Project \rm programs.  The methodology employed
parallels that used in reducing and analyzing the 18 galaxies observed as part
of the \it HST Key Project on the Extragalactic Distance Scale\rm.
This re-analysis has also benefited from the inclusion of additional archival
data not available at the time of original reduction, as well as the experience
and knowledge gained over the past five years in 
calibrating WF/PC and WFPC2 data.  Likewise, such cross-checking has
uncovered several minor discrepancies in some of the earlier published results.

The major result of our re-analysis is that, in the mean, the absolute distance
modulus for these seven
local Type Ia SNe calibrators is $0.12\pm 0.07$\,mag nearer (weighted), 
on average,
than those published originally by Tanvir \etal\ (1995) and Saha \etal\ 
(1994,1995,1996ab,1997,1999).  
There are several reasons for the systematic differences, including
differences in the photometry, Cepheid samples, weighting of
the data, and treatment of reddening.

Using these new distances, we have re-calculated the relevant zero points to
the Hubble Diagrams used by Suntzeff \etal\ (1999) and Saha \etal\ (1997), in
order to quantify their effect upon the predicted value of the Hubble constant.
Properly weighting each galaxy and SN by its associated uncertainties shows
that any previous empirical calibration tied to these
local Type Ia SNe calibrators will have underestimated H$_\circ$ by
$6-8$\%.

Formally, our favored \ho, based upon Type Ia SNe, is H$_\circ=68\pm 
2\,({\rm random})\pm 5\,({\rm systematic})$
\hounits.  This result assumes the decline rate-peak luminosity relationship of
Suntzeff \etal\ (1999) and no metallicity dependence of the underlying Cepheid
PL relationship.  If the Kennicutt \etal\ (1998) empirical 
metallicity dependence is assumed, the predicted H$_\circ$ is reduced to
H$_\circ=65\pm 2\,({\rm random})\pm 5\,({\rm systematic})$ \hounits.  
Our Type Ia SNe-based H$_\circ$ determination
agrees, to within the uncertainties,  with the \tf (Sakai \etal\ 1999) and
surface brightness
fluctuation (Ferrarese \etal\ 1999) results.

\acknowledgments

The work presented in this paper is based on observations with the NASA/ESA
Hubble Space Telescope, obtained by the Space Telescope Science Institute,
which is operated by AURA, Inc. under NASA contract No. 5-26555.
The continued assistance of the NASA and STScI support staff, and in particular
our program coordinator, Doug Van Orsow, is gratefully acknowledged.
Support for this work was provided by NASA through grant GO-2227-87A from
STScI. SMGH and PBS are grateful to NATO for travel support via a Collaborative
Research Grant (960178).   We wish to thank Nick Suntzeff and Brian Schmidt for
their input, and say a special word of thanks to the
referee, Mark Phillips, for a most thorough and helpful analysis of the paper.

\clearpage

\appendix{Appendix A: The Cepheid Sample}

Tables \ref{tbl:n4639_ceph} to \ref{tbl:n4496a_ceph} include basic positional
information, along with the Cepheid mean magnitudes
and periods, derived as
part of our ALLFRAME$+$TRIAL re-analysis of the Type Ia SNe-host galaxy
\it HST \rm archive data.  Stetson's (1998) photometric calibration was
employed, although transformation to the Hill \etal\ (1998) scale can be
accomplished (in the mean) by adding $+$0.02\,mag and $+$0.04\,mag,
respectively, to the Cepheid mean $V$- and $I$-band magnitudes.
Epoch-by-epoch photometry, light curves, and finding
charts, for all the Cepheids contained herein, are
available through the \it
HST Key Project\rm's web page -- {\tt http://www.ipac.caltech.edu/H0kp/}.

\placetable{tbl:n4639_ceph}

\placetable{tbl:n4536_ceph}

\placetable{tbl:n3627_ceph}

\placetable{tbl:n3368_ceph}

\placetable{tbl:n5253_ceph}

\placetable{tbl:i4182_ceph}

\placetable{tbl:n4496a_ceph}

\clearpage

\figcaption[n4639.eps]{
Period-luminosity relations in the $V$ (top panel) and $I$ (bottom panel) bands,
based on the Stetson (1998)-calibrated ALLFRAME photometry.  The filled
circles represent the 17
high-quality NGC~4639 Cepheid candidates found by TRIAL (see Table
\ref{tbl:n4639_ceph}).  The solid lines are least
squares fits to this entire sample,
with the slope fixed to be that of the Madore \& Freedman (1991)
LMC PL-relations, while 
the dotted lines represent their corresponding 2$\sigma$ dispersion.
The inferred apparent distance moduli are then
$\mu_V=31.92\pm 0.08$ mag (internal) and $\mu_I=31.87\pm 0.06$ mag
(internal).
\label{fig:n4639}}

\figcaption[n4536.eps]{
Period-luminosity relations in the $V$ (top panel) and $I$ (bottom panel) bands,
based on the Stetson (1998)-calibrated ALLFRAME photometry.  The filled
circles represent the 39
high-quality NGC~4536 Cepheid candidates found by TRIAL (see Table
\ref{tbl:n4536_ceph}).  The solid lines are least
squares fits to the 27 P$>$20\,d candidates, 
with the slope fixed to be that of the Madore \& Freedman (1991)
LMC PL-relations, while 
the dotted lines represent their corresponding 2$\sigma$ dispersion.
The inferred apparent distance moduli are then
$\mu_V=31.20\pm 0.06$ mag (internal) and $\mu_I=31.10\pm 0.04$ mag
(internal).
\label{fig:n4536}}

\figcaption[n4536_mod_comp.eps]{
\it Upper Panel: \rm
Distribution of dereddened true moduli
for the 29 Cepheids employed by Saha \etal\ (1996a) in deriving the distance 
to NGC~4536; the previously published Saha \etal\ DoPHOT photometry and periods
were adopted above.  The 17 Cepheids on Chips 1, 3, and 4, are self-consistent
and lead to a mean true modulus of $\mu_\circ=30.76$; the Chip 2 Cepheids (12,
in total) appear to lie 0.66\,mag beyond the Chips 1, 3, and 4 Cepheids.
Ignoring this discrepancy, and averaging the Cepheids from all four chips,
leads to the $\mu_\circ=31.10$ reported by Saha et~al.  \it Lower Panel: \rm
As for the upper panel, but now showing the results of our ALLFRAME$+$TRIAL
analysis.  No discrepancy is found between the 11 Chip~2 Cepheids and the 16
Chips 1, 3, and 4 Cepheids.  Averaging the Cepheids on all four chips leads to
the $\mu_\circ({\rm ALL})=30.95$ reported in Table \ref{tbl:distances}.
\label{fig:n4536_mod_comp}}

\figcaption[n3627.eps]{
Period-luminosity relations in the $V$ (top panel) and $I$ (bottom panel) bands,
based on the Stetson (1998)-calibrated ALLFRAME photometry.  The filled
circles represent the 36
high-quality NGC~3627 Cepheid candidates found by TRIAL (see Table
\ref{tbl:n3627_ceph}).  The solid lines are least
squares fits to the 17 P$>$25\,d candidates, 
with the slope fixed to be that of the Madore \& Freedman (1991)
LMC PL-relations, while 
the dotted lines represent their corresponding 2$\sigma$ dispersion.
The inferred apparent distance moduli are then
$\mu_V=30.40\pm 0.08$ mag (internal) and $\mu_I=30.26\pm 0.07$ mag
(internal).
\label{fig:n3627}}

\figcaption[n3627_mod_comp.eps]{
Distribution of dereddened true moduli for the final adopted NGC~3627 
Cepheid samples of Saha \etal\ (1999) (upper panel) and the current analysis
(lower panel).  The solid symbols in both panels represent the five Cepheids in
common between the two samples; for these five, there is excellent agreement
in both the $V$- and $I$-band mean magnitudes, resulting in excellent agreement
in the dereddened moduli.  Of the 0.23\,mag difference in the mean
true modulus, 0.14\,mag 
is due to the four Cepheid outliers with $\mu_\circ>30.8$ in the
upper panel (identified by the open circles).  
While the median and mean of the sample shown in the lower panel
are identical, they differ by 0.19\,mag in the Saha \etal\ sample (see Section
\ref{n3627} for details).
\label{fig:n3627_mod_comp}}

\figcaption[n3368.eps]{
Period-luminosity relations in the $V$ (top panel) and $I$ (bottom panel) bands,
based on the Stetson (1998)-calibrated ALLFRAME photometry.  The filled
circles represent the 11
high-quality NGC~3368 Cepheid candidates found by TRIAL (see Table
\ref{tbl:n3368_ceph}).  The solid lines are least
squares fits to the 7 P$>$20\,d candidates, 
with the slope fixed to be that of the Madore \& Freedman (1991)
LMC PL-relations, while 
the dotted lines represent their corresponding 2$\sigma$ dispersion.
The inferred apparent distance moduli are then
$\mu_V=30.55\pm 0.10$ mag (internal) and $\mu_I=30.41\pm 0.08$ mag
(internal).
\label{fig:n3368}}

\figcaption[n5253.eps]{
Period-luminosity relations in the $V$ (top panel) and $I$ (bottom panel) bands,
based on the Stetson (1998)-calibrated ALLFRAME photometry.  The filled
circles represent the 7
high-quality NGC~5253 Cepheid candidates found by TRIAL (see Table
\ref{tbl:n5253_ceph}).  The solid lines are least
squares fits to this entire sample,
with the slope fixed to be that of the Madore \& Freedman (1991)
LMC PL-relations, while 
the dotted lines represent their corresponding 2$\sigma$ dispersion.
The inferred apparent distance moduli are then
$\mu_V=27.95\pm 0.10$ mag (internal) and $\mu_I=27.82\pm 0.08$ mag
(internal). 
\label{fig:n5253}}

\figcaption[i4182.eps]{
Period-luminosity relations in the $V$ (top panel) and $I$ (bottom panel) bands,
based on the Stetson (1998)-calibrated ALLFRAME photometry.  The filled
circles represent the 28
high-quality IC~4182 Cepheid candidates found by TRIAL (see Table
\ref{tbl:i4182_ceph}), in common with Saha \etal\ (1994).  
The solid lines are least
squares fits to this entire sample,
with the slope fixed to be that of the Madore \& Freedman (1991)
LMC PL-relations, while 
the dotted lines represent their corresponding 2$\sigma$ dispersion.
The inferred apparent distance moduli are then
$\mu_V=28.28\pm 0.05$ mag (internal) and $\mu_I=28.31\pm 0.04$ mag
(internal). 
\label{fig:i4182}}

\figcaption[n4496a.eps]{
Period-luminosity relations in the $V$ (top panel) and $I$ (bottom panel) bands,
based on the Stetson (1998)-calibrated ALLFRAME photometry.  The filled
circles represent the 94
high-quality NGC~4496A Cepheid candidates found by TRIAL (see Table
\ref{tbl:n4496a_ceph}).  The solid lines are least
squares fits to the 51 P$>$25\,d candidates, 
with the slope fixed to be that of the Madore \& Freedman (1991)
LMC PL-relations, while 
the dotted lines represent their corresponding 2$\sigma$ dispersion.
The inferred apparent distance moduli are then
$\mu_V=31.12\pm 0.05$ mag (internal) and $\mu_I=31.08\pm 0.04$ mag
(internal).
\label{fig:n4496a}}

\clearpage

\begin{deluxetable}{lcll}
\footnotesize
\tablecaption{HST-Observed Type Ia Supernovae Host Galaxies
\label{tbl:galaxies}}
\tablewidth{0pt}
\tablehead{
\colhead{Galaxy} & 
\colhead{$12+\log({\rm O/H})$\tablenotemark{a}} & 
\colhead{Supernova} &
\colhead{HST Proposal ID\#}
}
\startdata
\multicolumn{4}{c}{\it High Quality SNe Light Curves} \nl
NGC~3982   & n/a           & SN~1998aq& \quad\qquad 8100\tablenotemark{b} \nl
NGC~4639   & 9.00$\pm$0.20 & SN~1990N & \quad\qquad 5981 \nl
NGC~4536   & 8.85$\pm$0.20 & SN~1981B & \quad\qquad 5427 \nl
NGC~3627   & 9.25$\pm$0.20 & SN~1989B & \quad\qquad 6549 \nl
NGC~3368   & 9.20$\pm$0.20 & SN~1998bu& \quad\qquad 5415 \nl
NGC~4527   & n/a           & SN~1991T & \quad\qquad 7504\tablenotemark{c} \nl\nl
\multicolumn{4}{c}{\it Intermediate Quality SNe Light Curves} \nl
NGC~5253   & 8.15$\pm$0.15 & SN~1972E & \quad\qquad 4277 \nl
NGC~1316   & n/a           & SN~1980N & \quad\qquad 7504,8100\tablenotemark{d} \nl
NGC~1316   & n/a           & SN~1981D & \quad\qquad 7504,8100\tablenotemark{d} \nl
IC~4182    & 8.40$\pm$0.20 & SN~1937C & \quad\qquad 2547 \nl\nl
\multicolumn{4}{c}{\it Low Quality SNe Light Curves} \nl
NGC~4496A  & 8.77$\pm$0.20 & SN~1960F & \quad\qquad 5427 \nl
NGC~4414   & 9.20$\pm$0.20 & SN~1974G & \quad\qquad 5397 \nl
NGC~5253   & 8.15$\pm$0.15 & SN~1895B & \quad\qquad 4277 \nl
\enddata
\tablenotetext{a}{H\,{\scriptsize II} region abundances from Kennicutt \etal\ 
(1999), on the calibration of Zaritsky \etal\ (1994).}
\tablenotetext{b}{Twelve epochs scheduled for Cycle 8.}
\tablenotetext{c}{Twelve epochs scheduled for Cycle 7.
Unfortunately NGC~4527 is host to the highly peculiar Type Ia 
supernova SN~1991T; the extreme nature of this event has been discussed
by Filippenko \& Riess (1999).}
\tablenotetext{d}{Single epoch feasibility study scheduled for Cycle 7; twelve
epochs tentatively scheduled for Cycle 8, pending the results of the
feasibility study.}
\end{deluxetable}

\clearpage

\begin{deluxetable}{lrrrrll}
\footnotesize
\tablecaption{Recent Determinations of H$_\circ$ with Type Ia Supernovae
\label{tbl:H0_Ia}}
\tablewidth{6.5in}
\tablehead{
\colhead{H$_\circ$\tablenotemark{a}} & 
\colhead{n$_{\rm Ia}$}  &
\colhead{n$_{\rm Ia}$}  &
\colhead{n$_{\rm Ia}$}  &
\colhead{n$_{\rm cal}$}  &
\colhead{$<{\rm M}_B^{\rm max}>$\tablenotemark{b}}  &
\colhead{Reference\qquad\qquad\qquad} \\
\colhead{[km\,s$^{-1}$\,Mpc$^{-1}$]} & 
\colhead{[z$>$0.1]}  &
\colhead{[0.01$<$z$<$0.1]}  &
\colhead{[z$<$0.01]}  &
\colhead{} &
\colhead{} &
\colhead{}
}
\startdata
\multicolumn{7}{c}{\it Empirical} \nl
\qquad 68$\pm$2$\pm$5   &   1$\;\;\;\;\;$  &  34$\;\;\;\;\;\;\;\;\;\;$  &   0$\;\;\;\;\;\;$  &   6$\;\;\;$  & -19.54$\pm$0.06 & This Work          \nl
\qquad 64$\pm$5$\pm$5   &   0$\;\;\;\;\;$  &  40$\;\;\;\;\;\;\;\;\;\;$  &   2$\;\;\;\;\;\;$  &   4$\;\;\;$  & $\qquad\;\;\;$? & Jha \etal\ (1999)  \nl
\qquad 60$\pm$2         &   0$\;\;\;\;\;$  &  40$\;\;\;\;\;\;\;\;\;\;$  &  13$\;\;\;\;\;\;$  &   8$\;\;\;$  & -19.49$\pm$0.07 & Saha \etal\ (1999) \nl
\qquad 63$\pm$2$\pm$4   &   1$\;\;\;\;\;$  &  39$\;\;\;\;\;\;\;\;\;\;$  &   0$\;\;\;\;\;\;$  &   6$\;\;\;$  & -19.47$\pm$0.03                           & Phillips \etal\ (1999) \nl
\qquad 64$\pm$2$\pm$4   &   1$\;\;\;\;\;$  &  34$\;\;\;\;\;\;\;\;\;\;$  &   0$\;\;\;\;\;\;$  &   5$\;\;\;$  & -19.42$\pm$0.04\tablenotemark{c}          & Suntzeff \etal\ (1999) \nl
\qquad 65$\pm$1$\pm$6   &  11$\;\;\;\;\;$  &  25$\;\;\;\;\;\;\;\;\;\;$  &   1$\;\;\;\;\;\;$  &   3$\;\;\;$  & -19.45$\pm$0.01\tablenotemark{d}          & Riess \etal\ (1998) \nl
\qquad 60$\pm$6   &   1$\;\;\;\;\;$  &  28$\;\;\;\;\;\;\;\;\;\;$  &   0$\;\;\;\;\;\;$  &   7$\;\;\;$  & -19.48$\pm$0.07 & Tripp (1998) \nl
\qquad 50$\pm$3$\pm$5   &   0$\;\;\;\;\;$  &  20$\;\;\;\;\;\;\;\;\;\;$  &  37$\;\;\;\;\;\;$  &   4$\;\;\;$  & -19.65$\pm$0.09 & Lanoix (1998) \nl
\qquad 58$\pm$3$\pm$8   &   0$\;\;\;\;\;$  &  40$\;\;\;\;\;\;\;\;\;\;$  &  15$\;\;\;\;\;\;$  &   7$\;\;\;$  & -19.52$\pm$0.07 & Saha \etal\ (1997) \nl
\qquad 60$\pm$5   &   7$\;\;\;\;\;$  &  25$\;\;\;\;\;\;\;\;\;\;$  &   0$\;\;\;\;\;\;$  &   7$\;\;\;$  & -19.48$\pm$0.07 & Tripp (1997) \nl
\qquad 63$\pm$3$\pm$3   &   1$\;\;\;\;\;$  &  28$\;\;\;\;\;\;\;\;\;\;$  &   0$\;\;\;\;\;\;$  &   4$\;\;\;$  & -19.14$\pm$0.03\tablenotemark{e} & Hamuy \etal\ (1996) \nl
\qquad 64$\pm$3$\pm$3   &   1$\;\;\;\;\;$  &  16$\;\;\;\;\;\;\;\;\;\;$  &   3$\;\;\;\;\;\;$  &   3$\;\;\;$  & -19.22$\pm$0.09\tablenotemark{f}          & Riess \etal\ (1996) \nl\nl
\multicolumn{7}{c}{\it Physical} \nl
\qquad 66$\pm$12  &   ?$\;\;\;\;\;$   &   ?$\;\;\;\;\;\;\;\;\;\;$  &   ?$\;\;\;\;\;\;$  & -$\;\;\;$  & $\qquad\;\;\;$?   & Iwamoto \& Nomoto (1998) \nl
\qquad 68$\pm$6$\pm$7   &   0$\;\;\;\;\;$  &   0$\;\;\;\;\;\;\;\;\;\;$  &   6$\;\;\;\;\;\;$  &   -$\;\;\;$  & -19.28$\pm$0.12\tablenotemark{g} & Ruiz-Lapuente (1996) \nl
\qquad 67$\pm$9   &   1$\;\;\;\;\;$  &   9$\;\;\;\;\;\;\;\;\;\;$  &  16$\;\;\;\;\;\;$  &   -$\;\;\;$  & -19.24$\pm$0.09\tablenotemark{h} & H\"oflich \& Khokhlov (1996) \nl
\enddata
\tablenotetext{a}{\scriptsize Uncertainties taken directly from the respective source
paper; when two uncertainties
are quoted, the first is ``internal'', while the second is ``external''.}
\tablenotetext{b}{\scriptsize Mean of the peak $B$-band magnitudes, for the relevant SNe
sample -- \ie, decline-rate-corrected magnitudes are  not 
reported here.  Quoted values for Phillips et~al., Suntzeff et~al., Riess
et~al., and Hamuy et~al. refer to their respective
distant sample of SNe, and not their nearby calibrating sample, while for this
work, and for that of Saha et~al, 
Lanoix, and Tripp, the
values refer to the mean of their local calibrating sample.}
\tablenotetext{c}{\scriptsize The weighted mean of M$_B^{\rm max}$ for the adopted
five calibrators (SNe~1937C, 1972E, 1981B, 1990N and 1998bu) is $<{\rm M}_B
^{\rm max}>=-19.68\pm 0.08$, 0.27\,mag brighter than that adopted in their
earlier study
(Hamuy \etal\ 1996), due to the treatment of foreground and host extinction,
and $\sim$27\% brighter than the mean for the distant sample (listed above).}
\tablenotetext{d}{\scriptsize The weighted mean $\Delta$ from Tables 5 and 10 of Riess
\etal\ (1998) is $-0.089$, which when applied to $<{\rm M}_B^{\rm max}>$,
from Table 2 of Riess \etal\ (1996) -- \ie, $-19.36$, yields the above mean of
$-19.45$.  n.b. this mean M$_B^{\rm max}$ corresponds to their high-$z$
sample, and  not  the nearby calibrating sample.  The nearby sample,
comprised of SNe~1972E, 1981B, and 1990N, are assumed to have $<{\rm M}_B
^{\rm max}>=-19.56\pm 0.10$, $\sim 11$\% brighter than the distant sample.}
\tablenotetext{e}{\scriptsize Weighted mean from Table 1 of Hamuy \etal\ (1996), adjusted
for their final \h0=63.1 \hounits.  For comparison, the weighted mean of their
four calibrators (SNe~1937C, 1972E, 1981B, and 1990N -- from their Table 2) is
$-19.41\pm 0.09$, $\sim 28$\% brighter than their distant sample.}
\tablenotetext{f}{\scriptsize The mean $\Delta$ from Table 3 of Riess \etal\ is $+0.136$, 
which
when applied to $<{\rm M}_B^{\rm max}>$ from their Table 2 (\ie, -19.36),
yields the above mean of -19.22.  n.b. this mean M$_B^{\rm max}$
corresponds to their high-$z$ sample, and  not  the nearby calibrating
sample.  The nearby sample, comprised of SNe~1972E, 1981B, and 1990N, are
assumed to have $<{\rm M}_B^{\rm max}>=-19.56\pm 0.10$, $\sim 37$\%
brighter than the distant sample.}
\tablenotetext{g}{\scriptsize Weighted mean of the 
6 SNe listed in Table 1 of Ruiz-Lapuente.}
\tablenotetext{h}{\scriptsize Unweighted mean of the first acceptable model listed in
column 8 of H\"oflich \& Khokhlov's Table 3.}
\end{deluxetable}

\clearpage

\hoffset=5.0mm
\begin{deluxetable}{rcccc}
\footnotesize
\tablecaption{Comparison of ALLFRAME and DoPHOT Cepheid Mean 
Magnitudes\tablenotemark{a}
\label{tbl:ceph_comp}}
\tablewidth{0pt}
\tablehead{
\colhead{\# Stars} & \colhead{$\Delta$V\tablenotemark{b}} &
\colhead{$\sigma_{\Delta V}$} &
\colhead{$\Delta$I\tablenotemark{b}} & \colhead{$\sigma_{\Delta I}$}
}
\startdata
\multicolumn{5}{c}{\it NGC~4639 Cepheids}\nl
 7$\;\;\;\;$ & $-$0.015 & 0.030 & $-$0.064 & 0.060 \nl\nl
\multicolumn{5}{c}{\it NGC~4536 Cepheids\tablenotemark{c}}\nl
16$\;\;\;\;$ & $-$0.113 & 0.018 & $-$0.006 & 0.034 \nl\nl
\multicolumn{5}{c}{\it NGC~3627 Cepheids}\nl
18$\;\;\;\;$ & $-$0.011 & 0.036 & $+$0.062 & 0.032 \nl\nl
\multicolumn{5}{c}{\it NGC~3368 Cepheids}\nl
 7$\;\;\;\;$ & $-$0.076 & 0.054 & $-$0.122 & 0.036 \nl\nl
\multicolumn{5}{c}{\it NGC~5253 Cepheids}\nl
 5$\;\;\;\;$ & $+$0.028 & 0.093 & $+$0.022 & 0.147 \nl\nl
\multicolumn{5}{c}{\it IC~4182 Cepheids}\nl
27$\;\;\;\;$ & $-$0.038 & 0.021 & $-$0.119 & 0.039 \nl\nl
\multicolumn{5}{c}{\it NGC~4496A Cepheids}\nl
45$\;\;\;\;$ & $-$0.039 & 0.014 & $-$0.033 & 0.028 \nl
\enddata
\tablenotetext{a}{Saha \etal\ (1994,1996a,1997,1999) and Tanvir \etal\ (1995)
do not report photometry for reference stars, so a direct comparison of 
photometry is possible only for the Cepheids in common.}
\tablenotetext{b}{$\Delta\equiv$ALLFRAME-DoPHOT.}
\tablenotetext{c}{The comparison shown refers to Chip 1,3, and 4 Cepheids only
(see discussion in Section \ref{n4536} for justification).  Blindly accepting
the full set of 28 overlapping Cepheids, regardless of chip, yields
$\Delta V=-0.036\pm 0.020$ and $\Delta I=-0.054\pm 0.024$.}
\end{deluxetable}

\clearpage

\begin{deluxetable}{lllll}
\footnotesize
\tablecaption{Comparison of Distance Moduli to Type Ia Supernova-Host 
Galaxies
\label{tbl:distances}}
\tablewidth{6.5in}
\tablehead{
\colhead{Galaxy} & 
\colhead{$\mu_\circ({\rm ALL})$\tablenotemark{a}} &
\colhead{$\mu_\circ({\rm KP})$\tablenotemark{b}} &
\colhead{$\mu_\circ({\rm pub})$\tablenotemark{c}} &
\colhead{$\mu_\circ({\rm SS,H_\circ})$\tablenotemark{d}}
}
\startdata
NGC~4639   & $31.80\pm 0.09\pm 0.16$ & 
	     $31.92\pm 0.11\pm 0.16$ & 
	     $32.03\pm 0.22$ &
	     $32.03\pm 0.22$ \nl
NGC~4536   & $30.95\pm 0.07\pm 0.16$ & 
	     $30.76\pm 0.12\pm 0.16$\tablenotemark{e} & 
	     $31.10\pm 0.13$ &
	     $31.10\pm 0.05$\tablenotemark{f} \nl
NGC~3627   & $30.06\pm 0.17\pm 0.16$ &
	     $30.15\pm 0.08\pm 0.16$\tablenotemark{g} &
	     $30.22 \pm 0.12$ &
	     $30.28 \pm 0.15$ \nl
NGC~3368   & $30.20\pm 0.10\pm 0.16$ & 
	     $30.36\pm 0.09\pm 0.16$ & 
	     $30.32\pm 0.16$ &
	     \qquad n/a \nl
NGC~5253   & $27.61\pm 0.11\pm 0.16$ &
	     $28.08\pm 0.34\pm 0.16$\tablenotemark{h}  & 
	     $28.08\pm 0.2$ &
	     $28.00\pm 0.08$\tablenotemark{i} \nl
IC~4182    & $28.36\pm 0.08\pm 0.16$                  &
	     $28.31\pm 0.08\pm 0.16$\tablenotemark{j} & 
	     $28.36\pm 0.09$ &
	     $28.36\pm 0.12$\tablenotemark{k} \nl
NGC~4496A  & $31.02\pm 0.07\pm 0.16$ &
	     $30.99\pm 0.07\pm 0.16$\tablenotemark{l} & 
	     $31.03\pm 0.14$ &
	     $31.13\pm 0.10$\tablenotemark{m} \nl
NGC~4414   & $31.41\pm 0.10\pm 0.16$ & 
	     \qquad\quad n/a & 
	     $31.41\pm 0.17$ &
	     \qquad n/a \nl
\enddata
\tablenotetext{a}{\scriptsize Our final adopted absolute distance moduli
(adopting the Hill \etal\ 1998 photometric calibration),
derived from the uniform ALLFRAME$+$TRIAL re-analysis described in
this paper.  The two uncertainties correspond to R$_{\rm PL}$ (first) and 
S$_{\rm PL}$ (second) of Table \ref{tbl:error}.  R$_V=3.3$ was assumed.}
\tablenotetext{b}{\scriptsize Best estimate of the absolute distance
moduli,  but using the tabulated original photometry  (sources noted in
point $c$), in conjunction with the standard \it Key Project \rm
PL fitting methodology adopted throughout this series.  The uncertainties 
parallel those of note 'a'.  R$_V=3.3$ was assumed.}
\tablenotetext{c}{\scriptsize Quoted absolute distance moduli -- NGC~4639
(Saha \etal\ 1997); NGC~4536 (Saha \etal\ 1996a); NGC~3627 (Saha \etal\ 1999);
NGC~3368 (Tanvir \etal\ 1995); NGC~5253 (Saha \etal\ 1995);
IC~4182 (Saha \etal\ 1994); NGC~4496A (Saha \etal\ 1996b); NGC~4414 (Turner
\etal\ 1998).}
\tablenotetext{d}{\scriptsize Absolute distance moduli adopted by Sandage,
Saha, and collaborators in their two primary H$_\circ$ references -- Sandage
\etal\ (1996) and Saha \etal\ (1997).}
\tablenotetext{e}{\scriptsize Standard PL-fitting to the 17 high-quality
Cepheids in Chips 1$+$3$+$4 (each chip being consistent with one another)
yields the above quoted distance modulus.  Including the 12 high-quality
Cepheids from the discrepant Chip 2 (``discrepant'' in the sense that
$\mu_\circ^{\rm Chip 2}=31.42\pm 0.12$\,(random)), 
with their self-consistent sample of 17, yields $\mu_\circ=31.03\pm 
0.11$\,(random).}
\tablenotetext{f}{\scriptsize Sandage \etal\ (1996) and Saha \etal\ (1997) adopt
their earlier published
$\mu_\circ$ (from Saha \etal\ 1996a), for their H$_\circ$
analyses, but appear to have mistakenly
adopted the uncertainty associated with $\mu_V$ 
\ie, 0.05 mag), instead of that associated with $\mu_\circ$ (\ie, 0.13
mag).}
\tablenotetext{g}{\scriptsize Derived by eliminating the four outliers
(C2-V8,C2-V19,C4-V2,C4-V4) with $\mu_\circ>30.8$
from the final set of 25 Cepheids employed by Saha
\etal\ (1999), and taking the unweighted mean of the dereddened moduli.}
\tablenotetext{h}{\scriptsize Derived from the four Cepheids from Table 4 of
Saha \etal\ (1995) with unambiguous periods P$>$7\,d, after elimination
of outliers C3-V1
and C4-V4, and those with large $I$-band photometric uncertainties (C3-V4 and
C3-V5).}
\tablenotetext{i}{\scriptsize It appears that
Sandage \etal\ (1996) and Saha \etal\ (1997) adopted $\mu_\circ=28.00$ instead of
their earlier published $\mu_\circ=28.08$ -- see Section \ref{n5253}.}
\tablenotetext{j}{\scriptsize Shown is the result of the standard PL
fitting methodology to the $V$-band data of Saha \etal\ (1994).  The inferred 
reddening, when using the $I$-band data (E(V$-$I)=$-$0.11$\pm$0.04), is
unphysical, and would lead us to rejecting it in the $\mu_\circ$ derivation.}
\tablenotetext{k}{\scriptsize Because the standard PL fitting led to an
unphysical negative reddening (see previous note), Saha \etal\ (1994)
assumed that A$_V$=A$_I$=0.0,
and, further, that $\mu_\circ\equiv(\mu_V+\mu_I)/2$.}
\tablenotetext{l}{\scriptsize Saha \etal\ (1996b) adopt an incompleteness bias
period cut of P=18\,d, while we favor a cut at P=25\,d.  While apparent in both
sets of data, it is perhaps more evident in the upper panel of Figure
\ref{fig:n4496a}.}
\tablenotetext{m}{\scriptsize Sandage \etal\ (1996) and Saha \etal\ (1997) 
appear to have mistakenly replaced their earlier published (Saha \etal\ 1996b)
absolute modulus ($\mu_\circ=31.03$) with the associated apparent $V$-band modulus
($\mu_V=31.13$).}
\end{deluxetable}

\clearpage

\hoffset=-12.0mm
\begin{deluxetable}{llccccccc}
\scriptsize
\tablecaption{Adopted Photometry and Light Curve Characteristics for the
Nearby (Calibrating) Supernovae Sample\tablenotemark{a}
\label{tbl:supernovae}}
\tablewidth{0pt}
\tablehead{
\colhead{SN} & 
\colhead{Galaxy} & 
\colhead{B$_{\rm max}$} &
\colhead{V$_{\rm max}$} &
\colhead{I$_{\rm max}$} &
\colhead{$\Delta m_{15}$(B)$_{\rm obs}$} &
\colhead{M$_B^{\rm max}$(ALL)} &
\colhead{M$_V^{\rm max}$(ALL)} &
\colhead{M$_I^{\rm max}$(ALL)}
}
\startdata
1990N\tablenotemark{b} & NGC~4639         & 
     $12.76\pm 0.03$   & $12.70\pm 0.02$  & $12.94\pm 0.02$  & $1.07\pm 0.05$ & 
     $-19.51\pm 0.12$  & $-19.46\pm 0.11$ & $-19.07\pm 0.10$ \nl
1981B\tablenotemark{b} & NGC~4536         & 
     $12.03\pm 0.03$   & $11.93\pm 0.03$  &     n/a         & $1.10\pm 0.07$ & 
     $-19.49\pm 0.11$  & $-19.45\pm 0.10$ &     n/a         \nl
1989B\tablenotemark{c} & NGC~3627         & 
     $12.34\pm 0.05$   & $12.02\pm 0.05$  &     n/a         & $1.31\pm 0.07$ & 
     $-19.25\pm 0.21$  & $-19.19\pm 0.20$ &     n/a         \nl
1998bu\tablenotemark{b}& NGC~3368         & 
     $12.20\pm 0.03$   & $11.88\pm 0.03$  & $11.67\pm 0.05$  & $1.01\pm 0.05$ &
     $-19.51\pm 0.16$  & $-19.46\pm 0.14$ & $-19.21\pm 0.13$ \nl
1972E\tablenotemark{b} & NGC~5253\tablenotemark{d}         & 
     $\;\;8.49\pm 0.14$   & $\;\;8.49\pm 0.15$  & $\;\;8.80\pm 0.19$  & $0.87\pm 0.10$ &
     $-19.40\pm 0.20$  & $-19.33\pm 0.20$ & $-18.94\pm 0.22$ \nl
1937C\tablenotemark{b} & IC~4182          & 
     $\;\;8.80\pm 0.09$   & $\;\;8.82\pm 0.11$  &     n/a         & $0.87\pm 0.10$ &
     $-19.74\pm 0.14$  & $-19.68\pm 0.15$ &     n/a         \nl
1960F\tablenotemark{d} & NGC~4496A        & 
     $11.77\pm 0.07$   & $11.51\pm 0.18$  &     n/a         & $1.06\pm 0.08$ &
     $-19.30\pm 0.11$  & $-19.55\pm 0.19$ &     n/a         \nl
1974G\tablenotemark{e} & NGC~4414         & 
     $12.48\pm 0.05$   & $12.30\pm 0.05$  &     n/a         & $1.11\pm 0.06$ &
     $-19.65\pm 0.31$  & $-19.67\pm 0.25$ &     n/a         \nl\nl
\multicolumn{4}{l}{\quad Weighted Mean (All 8 SNe)}           
&&& $-19.47\pm 0.05$ & $-19.47\pm 0.05$ & $-19.10\pm 0.07$ \nl
\multicolumn{4}{l}{\quad Weighted Mean (Excluding SN~1960F and 1974G)}  
&&& $-19.51\pm 0.06$ & $-19.46\pm 0.05$ & $-19.10\pm 0.07$ \nl
\enddata
\tablenotetext{a}{A ratio of total-to-selective absorption of R$_V=3.1$ was
adopted, in order to retain consistency with the Hubble relations of Suntzeff 
\etal\ (1999).}
\tablenotetext{b}{Suntzeff \etal\ (1999, Table 5).}
\tablenotetext{c}{Wells \etal\ (1994), Tables 8 and 9.  
Corrected for a total reddening of E(B$-$V)=0.37$\pm$0.03, the peak magnitudes
are B$_{\rm max}=10.82\pm 0.13$ and V$_{\rm max}=10.84\pm 0.11$.}
\tablenotetext{d}{Schaefer (1996).}
\tablenotetext{e}{Schaefer (1998a), uncorrected for internal reddening.
Corrected for a reddening of E(B$-$V)=0.16$\pm$0.07, the peak magnitudes are
B$_{\rm max}=11.82\pm 0.29$ and V$_{\rm max}=11.80\pm 0.22$.}
\end{deluxetable}

\clearpage

\begin{deluxetable}{lllll}
\footnotesize
\tablecaption{Values of H$_\circ$: The Peak Luminosity-Decline Rate
Relationship\tablenotemark{a}
\label{tbl:hamuy}}
\tablewidth{0pt}
\tablehead{
\colhead{SN} & 
\colhead{Galaxy} & 
\colhead{H$_\circ$(B)} &
\colhead{H$_\circ$(V)} &
\colhead{H$_\circ$(I)}
}
\startdata
1990N                  & NGC~4639 & $67.6\pm\;\; 6.6\pm\;\; 4.9$  & $67.4\pm\;\; 6.2\pm\;\; 4.9$ & $68.0\pm\;\; 6.4\pm\;\; 5.0$     \nl
1981B                  & NGC~4536 & $67.7\pm\;\; 6.5\pm\;\; 4.9$  & $66.9\pm\;\; 6.1\pm\;\; 4.9$ & \quad$\qquad$    n/a             \nl
1989B                  & NGC~3627 & $68.2\pm\;\; 9.6\pm\;\; 5.0$  & $68.9\pm\;\; 9.1\pm\;\; 5.0$ & \quad$\qquad$    n/a             \nl
1998bu                 & NGC~3368 & $68.6\pm\;\; 7.5\pm\;\; 5.0$  & $68.0\pm\;\; 6.9\pm\;\; 5.0$ & $64.6\pm\;\; 6.6\pm\;\; 4.7$     \nl
1972E                  & NGC~5253 & $74.6\pm    10.2\pm\;\; 5.4$  & $75.0\pm\;\; 9.9\pm\;\; 5.5$ & $78.3\pm    10.8\pm\;\; 5.7$     \nl
1937C                  & IC~4182  & $63.8\pm\;\; 7.8\pm\;\; 4.6$  & $64.0\pm\;\; 7.5\pm\;\; 4.7$ & \quad$\qquad$    n/a             \nl
1960F\tablenotemark{b} & NGC~4496A& $75.3\pm\;\; 7.3\pm\;\; 5.5$  & $65.2\pm\;\; 7.9\pm\;\; 4.7$ & \quad$\qquad$    n/a             \nl
1974G\tablenotemark{c} & NGC~4414 & $62.5\pm    10.3\pm\;\; 4.6$  & $60.2\pm\;\; 8.4\pm\;\; 4.4$ & \quad$\qquad$    n/a             \nl\nl
\multicolumn{2}{l}{\quad Weighted Mean (All 8 SNe)}             &
$68.6\pm\;\; 2.8\pm\;\; 5.0$  &   $66.7\pm\;\; 2.6\pm\;\; 4.9$  &   $68.2\pm\;\; 4.2\pm\;\; 5.1$  \nl
\multicolumn{2}{l}{\quad Weighted Mean (Excluding SN~1960F and 1974G)} & 
$67.9\pm\;\; 3.1\pm\;\; 5.0$  &   $67.7\pm\;\; 3.0\pm\;\; 5.0$  &   $68.2\pm\;\; 4.2\pm\;\; 5.1$  \nl
\enddata
\tablenotetext{a}{After Table 6 of Suntzeff \etal\ (1999).  R$_V=3.1$ adopted,
to retain consistency with Suntzeff et~al.}
\tablenotetext{b}{Light curve quality suspect; adopted
reddening correction  highly  uncertain.}
\tablenotetext{c}{Light curve quality low; Sandage \etal\ 
(1996) and Saha \etal\ (1997) do not employ this SN in their H$_\circ$
analyses.}
\end{deluxetable}

\clearpage

\hoffset=2.0mm
\begin{deluxetable}{rll}
\footnotesize
\tablecaption{Sample Error Budget: H$_\circ$(V) for SN~1981B in NGC~4536\tablenotemark{a}
\label{tbl:error}}
\tablewidth{0pt}
\tablehead{
\colhead{            } &
\colhead{Error Source} & 
\colhead{Error (mag) }
}
\startdata
{\bf I.} & {\bf ERRORS ON THE CEPHEID DISTANCE SCALE} & \nl
a & LMC True Modulus                                         & $\qquad\pm 0.13$ \nl
b & LMC PL zero point                                          & $\qquad\pm 0.02$ \nl
S1& LMC PL Systematic Uncertainty (=a+b)                     & $\qquad\pm 0.13$ \nl\nl

c & HST $V$-band Zero Point                                    & $\qquad\pm 0.03$ \nl
d & HST $I$-band Zero Point                                    & $\qquad\pm 0.03$ \nl
S2& Photometry Systematic Uncertainty ($=\sqrt{(1.47*{\rm c})^2+(2.47*{\rm d})^2}$)   & $\qquad\pm 0.09$ \nl\nl

R1& Photometry Random Uncertainty (ALLFRAME vs DoPHOT)       & $\qquad\pm 0.05$ \nl
R2& Extinction Treatment (uncertainty in R$_V$)        & $\qquad\pm 0.02$ \nl
R3& Dereddened PL fit                                        & $\qquad\pm 0.04$ \nl\nl
R$_{\rm PL}$& Total Random (=R1+R2+R3)                      & $\qquad\pm 0.07$ \nl
S$_{\rm PL}$& Total Systematic (=S1+S2)                     & $\qquad\pm 0.16$ \nl\nl

{\bf II.} & {\bf ERRORS ON THE SN DISTANCE SCALE} & \nl
e & Apparent $V$ Magnitude                                     & $\qquad\pm 0.03$ \nl
f & SN Line-of-Sight Reddening ($={\rm R}_V*\sigma_{E(\bmv)_t}$)& $\qquad\pm 0.06$ \nl
g & Distance Modulus (=R$_{\rm PL}$)                         & $\qquad\pm 0.07$ \nl
h & Absolute $V$ Magnitude (=e+f+g)                            & $\qquad\pm 0.10$ \nl\nl

i & M$_V^{\rm max}$ vs $\Delta m_{15}{(B)_{t}}$ Slope ($=0.396*[\Delta 
                                                        m_{15}(B)_{t}-1.1]$)
                                                             & $\qquad\pm 0.01$ \nl
j & M$_V^{\rm max}$ vs $\Delta m_{15}(B)_{t}$ Curvature ($=0.742*
                                                        [\Delta m_{15}
                                                        (B)_{t}-1.1]^2$)
                                                             & $\qquad\pm 0.00$ \nl
k & $\Delta m_{15}(B)_{t}$ ($=\sigma_{\Delta m_{15}(B)_{t}}[0.672+2*0.633*(\Delta m_{15}(B)_{t}-
                                                        1.1)]^2$)
                                                             & $\qquad\pm 0.03$ \nl
l & Corrected Absolute $V$ Magnitude (=h+i+j+k)                & $\qquad\pm 0.11$ \nl\nl

m & Hubble Diagram Zero Point                                & $\qquad\pm 0.04$\nl
n & Hubble Diagram Scatter                                   & $\qquad\pm 0.16$ \nl\nl
  & Total Random (=l+m+n)                                    & $\qquad\pm 0.20$ \nl
  &                                                     & =6.1\,\hounits  \nl
  & Total Systematic (=S$_{\rm PL}$)                         & $\qquad\pm 0.16$ \nl
  &                                                     & =4.9\,\hounits  \nl
\enddata
\tablenotetext{a}{After Table 4 of Hamuy \etal\ (1996).}
\end{deluxetable}

\clearpage

\begin{deluxetable}{lcccccc}
\footnotesize
\tablecaption{Summarizing the ALLFRAME$+$TRIAL H$_\circ$ Re-analysis
\label{tbl:H0_bkg}}
\tablewidth{0pt}
\tablehead{
\colhead{H$_\circ$\tablenotemark{a}} & 
\colhead{M$_{BVI}^{\rm max}$--$\Delta m_{15}(B)$\tablenotemark{b}} &
\colhead{n$_{\rm cal}$\tablenotemark{c}}  &
\colhead{Hubble\tablenotemark{d}} &
\colhead{$\mu_\circ$\tablenotemark{e}} &
\colhead{PL-[O/H]\tablenotemark{f}} &
\colhead{Reference\tablenotemark{g}} \\
\colhead{} &
\colhead{Correction} &
\colhead{} &
\colhead{Diagram} &
\colhead{Source} &
\colhead{Correction} &
\colhead{Frame}
}
\startdata
$67.9\pm 1.9\pm 5.0$                    & S99  & 6 & S99 & G99 & none & CMB \nl
$67.7\pm 1.7\pm 5.0$                    & S99  & 8 & S99 & G99 & none & CMB \nl
$64.9\pm 1.8\pm 4.9$                    & S99  & 6 & S99 & G99 & K98  & CMB \nl
$69.3\pm 3.0\pm 5.1$                    & S99  & 6 & S99 & G99 & none & FF  \nl
$62.7\pm 1.4\pm 4.6$\tablenotemark{h}   & none & 6 & S97 & G99 & none & CMB \nl
$65.5\pm 1.5\pm 4.8$\tablenotemark{h}   & S97  & 6 & S97 & G99 & none & CMB \nl
$61.5\pm 1.4\pm 4.5$\tablenotemark{h,i} & none & 6 & S97 & S/T & none & CMB \nl
\enddata
\tablenotetext{a}{Quoted uncertainties are ``random'' and ``systematic'',
respectively.  R$_V=3.1$ assumed for all entries employing the `S99' Hubble
Diagram, to retain consistency with Suntzeff \etal\ (1999).}
\tablenotetext{b}{Peak luminosity--light curve shape dependence employed:
S99$\equiv$Suntzeff \etal\ (1999); S97$\equiv$Saha \etal\ (1997).}
\tablenotetext{c}{Number of local Type Ia SNe calibrators employed: 
6 (SN~1990N, 1981B, 1989B, 1998bu, 1972E,
1937C); 8 (SN~1990N, 1981B, 1989B, 1998bu, 1972E, 1937C, 1960F, 1974G).}
\tablenotetext{d}{Hubble diagram SNe sample employed: S99$\equiv$Suntzeff
\etal\ 1999 (\ct Survey - Hamuy \etal\ 1996 $+$ CfA Survey 
- Riess \etal\ 1998); S97$\equiv$Saha \etal\ 1997 (\ct Survey $+$ Asiago Survey
- Patat 1995).}
\tablenotetext{e}{Source of calibrating galaxies' true moduli: G99$\equiv$this
work; S/T$\equiv$Saha \etal\ (1994,1995,1996ab,1997,1999) and Tanvir \etal\
(1995).}
\tablenotetext{f}{Cepheid PL-metallicity dependence employed:
K98$\equiv$Kennicutt \etal\ (1998).}
\tablenotetext{g}{CMB$\equiv$Cosmic Microwave Background; FF$\equiv$Flow Field
(Mould \etal\ 1999).}
\tablenotetext{h}{SNe photometry taken from Saha \etal\ (1997), and not Suntzeff
\etal\ (1999), although the differences are negligible.}
\tablenotetext{i}{Our replication of the results of Saha \etal\ (1997), using
galaxy distances as derived from their original photometry 
(column 3 of our Table \ref{tbl:distances}) -- in contrast, Saha \etal\ found
H$_\circ=57.6\pm 1.5\,({\rm random})$.}
\end{deluxetable}

\clearpage

\tablenum{A1}
\begin{deluxetable}{ccrrccc}
\footnotesize
\tablecaption{Cepheids Detected in NGC~4639 -- Properties
\label{tbl:n4639_ceph}}
\tablewidth{0pt}
\tablehead{
\colhead{ID} & \colhead{Chip} & \colhead{X} & \colhead{Y} &
\colhead{V (ALLFRAME)}  &
\colhead{I (ALLFRAME)}  &
\colhead{P (TRIAL)}
}
\startdata
C01 & 1 & 597.4 & 647.0 & $26.15\pm 0.03$ & $25.23\pm 0.06$ & $34.24\pm 0.53$\nl
C02 & 2 & 455.5 & 295.0 & $26.82\pm 0.06$ & $25.77\pm 0.08$ & $26.59\pm 0.68$\nl
C03 & 2 & 472.2 & 324.0 & $25.83\pm 0.02$ & $24.82\pm 0.05$ & $54.82\pm 1.25$\nl
C04 & 2 & 226.9 & 620.2 & $26.49\pm 0.03$ & $25.62\pm 0.07$ & $33.77\pm 1.31$\nl
C05 & 2 & 329.0 & 321.1 & $26.23\pm 0.04$ & $25.09\pm 0.06$ & $42.21\pm 0.91$\nl
C06 & 2 & 155.1 & 529.9 & $25.90\pm 0.03$ & $24.96\pm 0.08$ & $39.25\pm 0.69$\nl
C07 & 2 & 437.8 & 167.9 & $26.28\pm 0.04$ & $25.42\pm 0.08$ & $27.55\pm 0.54$\nl
C08 & 2 &  92.3 & 381.8 & $25.16\pm 0.02$ & $24.33\pm 0.05$ & $59.52\pm 1.59$\nl
C09 & 3 & 518.2 & 548.0 & $25.64\pm 0.03$ & $24.77\pm 0.06$ & $39.42\pm 0.93$\nl
C10 & 3 & 361.2 & 557.0 & $26.27\pm 0.04$ & $25.58\pm 0.08$ & $31.84\pm 1.13$\nl
C11 & 3 & 551.4 & 575.6 & $26.43\pm 0.03$ & $25.24\pm 0.04$ & $37.28\pm 0.32$\nl
C12 & 3 & 780.0 & 584.8 & $26.11\pm 0.04$ & $25.12\pm 0.07$ & $47.21\pm 1.92$\nl
C13 & 3 & 488.6 & 590.1 & $25.59\pm 0.02$ & $24.46\pm 0.05$ & $56.31\pm 0.88$\nl
C14 & 3 & 528.9 & 574.7 & $26.33\pm 0.03$ & $25.28\pm 0.05$ & $52.16\pm 0.80$\nl
C15 & 3 & 256.0 & 614.9 & $25.84\pm 0.02$ & $25.16\pm 0.04$ & $32.41\pm 0.70$\nl
C16 & 4 & 554.7 &  83.8 & $26.69\pm 0.05$ & $25.73\pm 0.10$ & $21.81\pm 0.82$\nl
C17 & 4 & 559.5 &  76.6 & $26.10\pm 0.02$ & $25.03\pm 0.05$ & $51.11\pm 1.27$\nl
\enddata
\end{deluxetable}

\clearpage

\tablenum{A2}
\begin{deluxetable}{ccrrccc}
\footnotesize
\tablecaption{Cepheids Detected in NGC~4536 -- Properties
\label{tbl:n4536_ceph}}
\tablewidth{0pt}
\tablehead{
\colhead{ID} & \colhead{Chip} & \colhead{X} & \colhead{Y} &
\colhead{V (ALLFRAME)}  &
\colhead{I (ALLFRAME)}  &
\colhead{P (TRIAL)}
}
\startdata
C01 & 1 & 210.3 & 155.6 & $26.40\pm 0.03$ & $25.62\pm 0.07$ & $15.13\pm 0.13$\nl
C02 & 1 & 228.6 & 291.5 & $25.67\pm 0.02$ & $24.63\pm 0.04$ & $31.59\pm 0.17$\nl
C03 & 1 & 106.4 & 375.6 & $25.94\pm 0.03$ & $25.21\pm 0.10$ & $20.98\pm 0.35$\nl
C04 & 1 & 252.4 & 651.5 & $25.72\pm 0.03$ & $24.55\pm 0.04$ & $34.63\pm 0.41$\nl
C05 & 1 &  82.4 & 661.7 & $26.01\pm 0.04$ & $24.92\pm 0.06$ & $30.25\pm 0.49$\nl
C06 & 1 & 375.9 & 471.8 & $25.29\pm 0.02$ & $24.41\pm 0.04$ & $37.45\pm 0.40$\nl
C07 & 1 & 713.2 & 639.1 & $26.31\pm 0.04$ & $25.12\pm 0.06$ & $23.23\pm 0.22$\nl
C08 & 2 &  89.4 & 202.4 & $26.15\pm 0.03$ & $25.21\pm 0.06$ & $27.61\pm 0.78$\nl
C09 & 2 & 686.7 & 157.7 & $25.90\pm 0.04$ & $24.94\pm 0.05$ & $30.06\pm 0.46$\nl
C10 & 2 & 423.7 & 186.6 & $26.07\pm 0.03$ & $25.21\pm 0.05$ & $15.00\pm 0.15$\nl
C11 & 2 & 291.8 & 255.8 & $25.33\pm 0.02$ & $24.40\pm 0.03$ & $33.67\pm 0.74$\nl
C12 & 2 & 593.7 & 328.2 & $25.81\pm 0.03$ & $24.89\pm 0.05$ & $30.45\pm 0.45$\nl
C13 & 2 & 399.3 & 405.5 & $26.10\pm 0.04$ & $25.43\pm 0.06$ & $18.54\pm 0.14$\nl
C14 & 2 & 342.9 & 454.0 & $26.02\pm 0.03$ & $24.97\pm 0.04$ & $19.79\pm 0.35$\nl
C15 & 2 & 402.4 & 474.4 & $25.63\pm 0.02$ & $24.62\pm 0.03$ & $42.81\pm 0.63$\nl
C16 & 2 & 144.7 & 523.5 & $24.80\pm 0.01$ & $23.75\pm 0.02$ & $55.17\pm 0.96$\nl
C17 & 2 & 284.3 & 572.6 & $25.17\pm 0.02$ & $24.04\pm 0.03$ & $49.71\pm 0.59$\nl
C18 & 2 &  73.6 &  86.4 & $26.02\pm 0.06$ & $24.93\pm 0.07$ & $28.16\pm 0.62$\nl
C19 & 2 & 726.0 & 133.8 & $25.39\pm 0.02$ & $24.46\pm 0.03$ & $33.28\pm 0.29$\nl
C20 & 2 & 535.2 & 340.3 & $26.24\pm 0.03$ & $25.10\pm 0.05$ & $29.49\pm 0.56$\nl
C21 & 2 & 195.5 & 164.9 & $26.49\pm 0.04$ & $25.53\pm 0.07$ & $18.85\pm 0.25$\nl
C22 & 2 & 373.4 & 127.8 & $26.38\pm 0.05$ & $25.25\pm 0.06$ & $21.54\pm 0.16$\nl
C23 & 2 & 372.6 & 344.9 & $26.29\pm 0.04$ & $25.25\pm 0.06$ & $19.74\pm 0.21$\nl
C24 & 3 & 196.2 & 252.6 & $25.96\pm 0.03$ & $25.01\pm 0.05$ & $22.38\pm 0.22$\nl
C25 & 3 &  91.9 & 489.5 & $26.39\pm 0.05$ & $25.38\pm 0.06$ & $16.78\pm 0.17$\nl
C26 & 3 & 483.4 & 543.1 & $25.46\pm 0.02$ & $24.45\pm 0.03$ & $38.57\pm 0.66$\nl
C27 & 3 & 311.5 & 574.3 & $24.31\pm 0.01$ & $23.32\pm 0.02$ & $55.24\pm 0.88$\nl
C28 & 3 & 621.3 & 646.8 & $25.89\pm 0.02$ & $25.17\pm 0.05$ & $19.08\pm 0.15$\nl
C29 & 3 & 222.7 & 677.0 & $25.45\pm 0.03$ & $24.76\pm 0.04$ & $24.55\pm 0.32$\nl
C30 & 3 & 359.8 & 703.0 & $26.24\pm 0.04$ & $25.37\pm 0.06$ & $14.29\pm 0.21$\nl
C31 & 3 & 266.8 & 474.4 & $24.75\pm 0.02$ & $23.74\pm 0.03$ & $53.17\pm 0.79$\nl
C32 & 3 & 279.0 & 216.0 & $25.21\pm 0.02$ & $24.27\pm 0.03$ & $38.01\pm 0.49$\nl
C33 & 4 & 660.9 & 137.5 & $25.27\pm 0.02$ & $24.21\pm 0.03$ & $42.99\pm 0.67$\nl
C34 & 4 & 536.6 & 574.9 & $26.11\pm 0.03$ & $25.42\pm 0.08$ & $17.10\pm 0.17$\nl
C35 & 4 & 268.5 & 766.8 & $25.54\pm 0.02$ & $24.78\pm 0.05$ & $24.16\pm 0.29$\nl
C36 & 4 & 500.8 & 392.0 & $25.71\pm 0.03$ & $24.93\pm 0.05$ & $23.38\pm 0.34$\nl
C37 & 4 & 781.0 & 440.9 & $26.38\pm 0.03$ & $25.61\pm 0.07$ & $17.72\pm 0.23$\nl
C38 & 4 & 393.0 & 352.2 & $25.37\pm 0.02$ & $24.24\pm 0.03$ & $51.49\pm 0.60$\nl
C39 & 4 & 422.2 &  98.5 & $26.34\pm 0.04$ & $25.55\pm 0.08$ & $18.22\pm 0.19$\nl
\enddata
\end{deluxetable}

\clearpage

\tablenum{A3}
\begin{deluxetable}{ccrrccc}
\footnotesize
\tablecaption{Cepheids Detected in NGC~3627 -- Properties
\label{tbl:n3627_ceph}}
\tablewidth{0pt}
\tablehead{
\colhead{ID} & \colhead{Chip} & \colhead{X} & \colhead{Y} &
\colhead{V (ALLFRAME)}  &
\colhead{I (ALLFRAME)}  &
\colhead{P (TRIAL)}
}
\startdata
C01 & 1 & 304.0 & 343.0 & $26.35\pm 0.03$ & $25.21\pm 0.04$ & $14.00\pm 0.12$\nl
C02 & 1 & 254.7 & 641.0 & $25.43\pm 0.02$ & $24.24\pm 0.04$ & $16.78\pm 0.18$\nl
C03 & 2 & 145.4 & 171.0 & $25.17\pm 0.03$ & $24.14\pm 0.07$ & $17.81\pm 0.33$\nl
C04 & 2 & 124.1 & 199.4 & $24.57\pm 0.02$ & $23.46\pm 0.02$ & $41.19\pm 0.84$\nl
C05 & 2 & 389.8 & 199.8 & $25.39\pm 0.04$ & $24.31\pm 0.04$ & $29.04\pm 0.71$\nl
C06 & 2 & 308.4 & 227.3 & $25.48\pm 0.03$ & $24.79\pm 0.08$ & $18.19\pm 0.17$\nl
C07 & 2 & 416.6 & 267.6 & $24.97\pm 0.02$ & $23.74\pm 0.03$ & $27.36\pm 0.50$\nl
C08 & 2 & 517.9 & 282.3 & $25.12\pm 0.02$ & $24.00\pm 0.03$ & $15.46\pm 0.12$\nl
C09 & 2 & 287.4 & 307.4 & $24.95\pm 0.03$ & $24.13\pm 0.05$ & $25.59\pm 0.55$\nl
C10 & 2 & 170.9 & 308.7 & $25.03\pm 0.03$ & $24.10\pm 0.05$ & $38.54\pm 1.08$\nl
C11 & 2 & 516.4 & 343.5 & $25.25\pm 0.04$ & $23.96\pm 0.04$ & $32.73\pm 0.85$\nl
C12 & 2 & 549.2 & 245.8 & $24.78\pm 0.02$ & $23.80\pm 0.03$ & $30.55\pm 0.36$\nl
C13 & 2 & 183.6 & 347.3 & $24.30\pm 0.03$ & $23.50\pm 0.03$ & $24.57\pm 0.47$\nl
C14 & 2 &  89.8 & 388.5 & $24.66\pm 0.03$ & $23.46\pm 0.05$ & $23.25\pm 0.40$\nl
C15 & 2 & 751.4 & 465.2 & $23.96\pm 0.02$ & $23.26\pm 0.03$ & $32.98\pm 1.04$\nl
C16 & 2 & 642.4 & 762.0 & $24.37\pm 0.02$ & $23.32\pm 0.03$ & $55.01\pm 1.88$\nl
C17 & 2 & 469.8 & 291.2 & $24.66\pm 0.02$ & $23.93\pm 0.03$ & $24.89\pm 0.41$\nl
C18 & 2 & 484.5 & 349.4 & $24.55\pm 0.01$ & $23.45\pm 0.02$ & $18.38\pm 0.23$\nl
C19 & 2 & 307.5 & 392.3 & $25.26\pm 0.03$ & $24.42\pm 0.07$ & $23.62\pm 0.28$\nl
C20 & 2 & 742.8 & 289.0 & $24.66\pm 0.02$ & $23.55\pm 0.03$ & $38.31\pm 0.74$\nl
C21 & 2 & 648.2 & 311.4 & $25.16\pm 0.02$ & $24.11\pm 0.03$ & $13.37\pm 0.21$\nl
C22 & 2 & 353.3 & 776.1 & $24.28\pm 0.02$ & $23.24\pm 0.02$ & $50.45\pm 1.88$\nl
C23 & 2 & 582.0 & 261.3 & $24.37\pm 0.02$ & $23.26\pm 0.02$ & $45.95\pm 0.44$\nl
C24 & 3 & 350.0 & 120.1 & $24.10\pm 0.02$ & $23.20\pm 0.03$ & $43.77\pm 0.85$\nl
C25 & 3 & 255.2 & 209.7 & $25.28\pm 0.03$ & $24.31\pm 0.05$ & $19.68\pm 0.25$\nl
C26 & 3 & 749.1 & 275.6 & $25.23\pm 0.03$ & $24.32\pm 0.05$ & $24.82\pm 0.34$\nl
C27 & 3 & 328.6 & 281.2 & $25.08\pm 0.02$ & $23.87\pm 0.02$ & $34.81\pm 0.68$\nl
C28 & 3 & 187.3 & 329.4 & $24.84\pm 0.03$ & $23.91\pm 0.05$ & $27.73\pm 0.72$\nl
C29 & 3 & 243.2 & 394.0 & $24.84\pm 0.03$ & $24.14\pm 0.04$ & $18.92\pm 0.23$\nl
C30 & 3 & 416.0 & 652.3 & $25.52\pm 0.03$ & $24.47\pm 0.04$ & $24.16\pm 0.22$\nl
C31 & 3 & 520.5 & 662.1 & $25.29\pm 0.03$ & $24.18\pm 0.04$ & $14.29\pm 0.14$\nl
C32 & 3 & 418.0 & 235.3 & $24.93\pm 0.04$ & $23.94\pm 0.05$ & $31.65\pm 1.75$\nl
C33 & 4 & 601.9 & 163.1 & $25.56\pm 0.04$ & $24.35\pm 0.05$ & $24.44\pm 0.20$\nl
C34 & 4 & 585.6 & 185.5 & $25.21\pm 0.03$ & $24.15\pm 0.04$ & $24.74\pm 0.33$\nl
C35 & 4 & 535.6 & 471.7 & $25.79\pm 0.04$ & $24.88\pm 0.05$ & $12.51\pm 0.10$\nl
C36 & 4 & 135.7 & 125.4 & $23.81\pm 0.01$ & $22.67\pm 0.02$ & $53.08\pm 0.99$\nl
\enddata
\end{deluxetable}

\clearpage

\tablenum{A4}
\begin{deluxetable}{crrccc}
\footnotesize
\tablecaption{Cepheids Detected in NGC~3368 -- Properties
\label{tbl:n3368_ceph}}
\tablewidth{0pt}
\tablehead{
\colhead{ID} & \colhead{X\tablenotemark{a}} & \colhead{Y\tablenotemark{a}} &
\colhead{V (ALLFRAME)}  &
\colhead{I (ALLFRAME)}  &
\colhead{P (TRIAL)}
}
\startdata
C01 & 723.8 & 575.5 & $25.75\pm 0.04$ & $24.72\pm 0.07$ & $12.44\pm 0.02$ \nl
C02 & 702.1 & 316.5 & $25.38\pm 0.04$ & $24.55\pm 0.05$ & $14.06\pm 0.02$ \nl
C03 &-689.0 & 266.5 & $25.83\pm 0.05$ & $24.94\pm 0.08$ & $15.05\pm 0.19$ \nl
C04 & 628.9 & 527.4 & $25.24\pm 0.03$ & $24.55\pm 0.08$ & $15.12\pm 0.04$ \nl
C05 &-328.9 &-470.8 & $25.15\pm 0.02$ & $24.32\pm 0.04$ & $21.22\pm 0.06$ \nl
C06 &-536.7 & 392.0 & $25.02\pm 0.03$ & $24.03\pm 0.03$ & $27.48\pm 0.06$ \nl
C07 & 764.6 & 380.5 & $24.83\pm 0.07$ & $23.78\pm 0.03$ & $28.57\pm 0.18$ \nl
C08 &-113.3 &-334.4 & $25.37\pm 0.03$ & $24.20\pm 0.04$ & $29.43\pm 0.10$ \nl
C09 & -94.8 & 656.5 & $25.13\pm 0.02$ & $24.11\pm 0.04$ & $30.39\pm 0.10$ \nl
C10 &-446.5 & 252.3 & $25.32\pm 0.03$ & $24.26\pm 0.04$ & $30.59\pm 0.13$ \nl
C11 & -42.4 & 690.9 & $24.67\pm 0.02$ & $23.70\pm 0.04$ & $43.45\pm 0.31$ \nl
\enddata
\tablenotetext{a}{Pixel coordinates are referenced with respect to $HST$
archive image {\tt u26g0101t2}.}
\end{deluxetable}

\clearpage

\tablenum{A5}
\begin{deluxetable}{crrccr}
\footnotesize
\tablecaption{Cepheids Detected in NGC~5253 -- Properties
\label{tbl:n5253_ceph}}
\tablewidth{0pt}
\tablehead{
\colhead{ID} & \colhead{X\tablenotemark{a}} & \colhead{Y\tablenotemark{a}} &
\colhead{V (ALLFRAME)}  &
\colhead{I (ALLFRAME)}  &
\colhead{P (TRIAL)}
}
\startdata
C01 & 336.1 & 118.8 & $24.25\pm 0.02$ & $23.35\pm 0.07$ & $ 6.54\pm 0.05$ \nl
C02 & 535.7 & 163.1 & $23.89\pm 0.02$ & $22.78\pm 0.03$ & $13.83\pm 0.01$ \nl
C03 & 738.8 & 386.8 & $23.43\pm 0.02$ & $22.74\pm 0.03$ & $11.99\pm 0.00$ \nl
C04 & 299.3 & 529.2 & $21.94\pm 0.01$ & $20.94\pm 0.02$ & $36.24\pm 0.05$ \nl
C05 &1124.0 & 549.7 & $24.33\pm 0.03$ & $23.48\pm 0.11$ & $ 5.54\pm 0.04$ \nl
C06 & 393.8 &-562.0 & $25.26\pm 0.04$ & $24.60\pm 0.12$ & $ 3.16\pm 0.01$ \nl
C07 & 964.1 & 433.6 & $23.71\pm 0.02$ & $22.86\pm 0.04$ & $10.60\pm 0.00$ \nl
\enddata
\tablenotetext{a}{Pixel coordinates are referenced with respect to $HST$
archive image {\tt u3760105t3}.}
\end{deluxetable}

\clearpage

\tablenum{A6}
\begin{deluxetable}{crrccr}
\footnotesize
\tablecaption{Cepheids Detected in IC~4182 -- Properties
\label{tbl:i4182_ceph}}
\tablewidth{0pt}
\tablehead{
\colhead{ID} & \colhead{X\tablenotemark{a}} & \colhead{Y\tablenotemark{a}} &
\colhead{V (ALLFRAME)}  &
\colhead{I (ALLFRAME)}  &
\colhead{P (TRIAL)}
}
\startdata
C01 &-200.5 & 151.8 & $22.35\pm 0.01$ & $21.36\pm 0.04$ & $58.75\pm 0.14$ \nl
C02 &-479.8 & 188.3 & $22.83\pm 0.01$ & $22.08\pm 0.07$ & $21.38\pm 0.19$ \nl
C03 &-503.5 & 305.1 & $24.00\pm 0.02$ & $22.95\pm 0.10$ & $12.87\pm 0.03$ \nl
C04 &  96.2 & 643.0 & $24.93\pm 0.03$ & $24.42\pm 0.26$ & $ 4.30\pm 0.03$ \nl
C05 &-481.3 & 338.5 & $24.56\pm 0.02$ & $23.75\pm 0.09$ & $10.67\pm 0.04$ \nl
C06 &-216.8 & 291.4 & $24.98\pm 0.04$ & $24.21\pm 0.16$ & $ 6.07\pm 0.01$ \nl
C07 &-372.7 &  29.9 & $22.25\pm 0.01$ & $21.51\pm 0.03$ & $42.03\pm 0.39$ \nl
C08 &-331.9 &  60.4 & $23.10\pm 0.01$ & $22.43\pm 0.05$ & $18.06\pm 0.02$ \nl
C09 &-259.2 &-141.8 & $22.53\pm 0.01$ & $21.76\pm 0.03$ & $35.75\pm 0.04$ \nl
C10 &-257.8 &  33.3 & $22.71\pm 0.01$ & $21.99\pm 0.05$ & $39.00\pm 0.04$ \nl
C11 &-653.5 & -56.8 & $23.10\pm 0.01$ & $22.21\pm 0.04$ & $26.51\pm 0.17$ \nl
C12 &-521.5 &-150.1 & $23.47\pm 0.01$ & $22.74\pm 0.06$ & $15.43\pm 0.01$ \nl
C13 &-455.6 &  63.2 & $23.37\pm 0.01$ & $22.48\pm 0.03$ & $22.72\pm 0.04$ \nl
C14 &-526.5 &-460.8 & $23.33\pm 0.01$ & $22.46\pm 0.05$ & $20.39\pm 0.06$ \nl
C15 &-240.0 &-145.1 & $24.77\pm 0.03$ & $24.33\pm 0.16$ & $ 4.36\pm 0.00$ \nl
C16 &-326.3 &-570.0 & $24.52\pm 0.03$ & $23.67\pm 0.11$ & $ 6.93\pm 0.02$ \nl
C17 &-645.1 &-449.4 & $24.57\pm 0.02$ & $23.99\pm 0.14$ & $ 7.49\pm 0.01$ \nl
C18 & -96.6 & -59.9 & $24.74\pm 0.02$ & $24.36\pm 0.21$ & $ 5.17\pm 0.01$ \nl
C19 & -79.3 &-550.7 & $24.77\pm 0.02$ & $24.15\pm 0.14$ & $ 5.83\pm 0.01$ \nl
C20 & 115.3 &   7.1 & $25.10\pm 0.04$ & $24.62\pm 0.31$ & $ 3.69\pm 0.04$ \nl
C21 & 156.6 & 101.4 & $21.93\pm 0.01$ & $21.28\pm 0.03$ & $41.07\pm 0.16$ \nl
C22 & 151.5 &-160.4 & $23.23\pm 0.01$ & $22.46\pm 0.04$ & $25.43\pm 0.19$ \nl
C23 & 139.1 &-175.8 & $24.02\pm 0.02$ & $23.51\pm 0.08$ & $ 9.05\pm 0.01$ \nl
C24 & 561.9 &-235.3 & $24.59\pm 0.02$ & $23.69\pm 0.09$ & $ 7.31\pm 0.01$ \nl
C25 & 594.5 &-129.0 & $24.49\pm 0.02$ & $23.77\pm 0.09$ & $ 7.05\pm 0.02$ \nl
C26 & 531.8 & 262.3 & $22.84\pm 0.01$ & $21.89\pm 0.03$ & $37.53\pm 0.06$ \nl
C27 & 360.3 & 191.7 & $24.38\pm 0.02$ & $23.65\pm 0.13$ & $ 6.52\pm 0.01$ \nl
C28 & 579.0 & 174.9 & $24.58\pm 0.02$ & $23.86\pm 0.11$ & $ 5.73\pm 0.04$ \nl
\enddata
\tablenotetext{a}{Pixel coordinates are referenced with respect to $HST$
archive image {\tt w18g0102t2}.}
\end{deluxetable}

\clearpage

\tablenum{A7}
\begin{deluxetable}{ccrrccc}
\footnotesize
\tablecaption{Cepheids Detected in NGC~4496A -- Properties
\label{tbl:n4496a_ceph}}
\tablewidth{0pt}
\tablehead{
\colhead{ID} & \colhead{Chip} & \colhead{X} & \colhead{Y} &
\colhead{V (ALLFRAME)}  &
\colhead{I (ALLFRAME)}  &
\colhead{P (TRIAL)}
}
\startdata
C01 & 1 & 289.2 & 110.5 & $26.17\pm 0.03$ & $25.38\pm 0.08$ & $19.62\pm 0.14$\nl
C02 & 1 & 779.8 & 465.3 & $26.16\pm 0.03$ & $25.20\pm 0.07$ & $22.55\pm 0.43$\nl
C03 & 1 & 134.7 & 696.6 & $25.55\pm 0.02$ & $24.63\pm 0.05$ & $32.86\pm 0.39$\nl
C04 & 1 & 343.8 & 279.7 & $26.80\pm 0.04$ & $25.82\pm 0.09$ & $15.66\pm 0.22$\nl
C05 & 1 & 266.3 & 379.2 & $26.02\pm 0.02$ & $25.07\pm 0.05$ & $28.79\pm 0.46$\nl
C06 & 1 & 241.8 & 481.2 & $26.19\pm 0.03$ & $25.25\pm 0.06$ & $17.05\pm 0.21$\nl
C07 & 1 & 663.2 & 524.5 & $25.43\pm 0.02$ & $24.43\pm 0.04$ & $29.11\pm 0.42$\nl
C08 & 1 & 364.4 & 535.7 & $26.27\pm 0.02$ & $25.50\pm 0.05$ & $18.80\pm 0.17$\nl
C09 & 1 & 680.4 & 546.0 & $26.15\pm 0.04$ & $25.42\pm 0.08$ & $15.93\pm 0.21$\nl
C10 & 1 & 659.2 & 564.8 & $24.73\pm 0.02$ & $23.96\pm 0.04$ & $46.92\pm 0.83$\nl
C11 & 1 & 765.4 & 318.6 & $24.95\pm 0.01$ & $24.26\pm 0.02$ & $32.41\pm 0.34$\nl
C12 & 1 & 424.9 & 415.1 & $26.34\pm 0.03$ & $25.42\pm 0.10$ & $19.69\pm 0.33$\nl
C13 & 1 & 693.9 & 288.7 & $24.97\pm 0.01$ & $23.89\pm 0.03$ & $68.29\pm 1.21$\nl
C14 & 2 & 285.7 & 164.8 & $24.43\pm 0.01$ & $23.42\pm 0.02$ & $63.75\pm 1.83$\nl
C15 & 2 & 554.7 & 178.0 & $26.53\pm 0.05$ & $25.45\pm 0.06$ & $22.18\pm 0.26$\nl
C16 & 2 & 735.7 & 279.0 & $26.32\pm 0.03$ & $25.30\pm 0.06$ & $17.95\pm 0.24$\nl
C17 & 2 & 286.6 & 280.3 & $24.48\pm 0.01$ & $23.65\pm 0.02$ & $51.49\pm 1.33$\nl
C18 & 2 & 261.8 & 336.5 & $25.67\pm 0.03$ & $24.57\pm 0.05$ & $42.18\pm 0.95$\nl
C19 & 2 & 435.6 & 356.9 & $26.15\pm 0.04$ & $25.38\pm 0.05$ & $21.17\pm 0.21$\nl
C20 & 2 & 389.2 & 375.5 & $26.56\pm 0.04$ & $25.44\pm 0.06$ & $13.97\pm 0.15$\nl
C21 & 2 & 514.6 & 377.4 & $26.24\pm 0.04$ & $25.35\pm 0.09$ & $23.28\pm 0.38$\nl
C22 & 2 & 361.4 & 395.8 & $26.28\pm 0.04$ & $25.34\pm 0.08$ & $16.13\pm 0.31$\nl
C23 & 2 & 171.1 & 426.1 & $24.93\pm 0.01$ & $23.96\pm 0.02$ & $45.17\pm 0.90$\nl
C24 & 2 & 473.3 & 437.1 & $25.27\pm 0.02$ & $24.26\pm 0.02$ & $52.07\pm 0.74$\nl
C25 & 2 & 102.0 & 468.0 & $26.45\pm 0.03$ & $25.36\pm 0.07$ & $22.82\pm 0.30$\nl
C26 & 2 & 162.1 & 494.8 & $25.66\pm 0.03$ & $24.87\pm 0.04$ & $28.60\pm 0.28$\nl
C27 & 2 & 365.1 & 497.0 & $25.22\pm 0.01$ & $24.25\pm 0.03$ & $39.12\pm 0.28$\nl
C28 & 2 & 228.6 & 557.9 & $25.32\pm 0.02$ & $24.18\pm 0.03$ & $60.74\pm 0.87$\nl
C29 & 2 & 521.8 & 571.5 & $26.21\pm 0.02$ & $25.16\pm 0.05$ & $21.81\pm 0.31$\nl
C30 & 2 & 161.6 & 591.6 & $26.50\pm 0.04$ & $25.41\pm 0.08$ & $14.58\pm 0.10$\nl
C31 & 2 & 391.7 & 676.5 & $25.42\pm 0.02$ & $24.58\pm 0.04$ & $28.06\pm 0.16$\nl
C32 & 2 & 622.8 & 752.0 & $26.39\pm 0.03$ & $25.58\pm 0.06$ & $14.91\pm 0.13$\nl
C33 & 2 & 422.5 & 219.6 & $24.78\pm 0.01$ & $23.82\pm 0.04$ & $47.72\pm 0.78$\nl
C34 & 2 & 450.9 & 269.1 & $26.32\pm 0.04$ & $25.51\pm 0.09$ & $17.26\pm 0.28$\nl
C35 & 2 & 444.9 & 297.6 & $25.37\pm 0.02$ & $24.23\pm 0.03$ & $51.68\pm 0.50$\nl
C36 & 2 & 561.9 & 420.9 & $26.18\pm 0.03$ & $25.28\pm 0.07$ & $15.89\pm 0.17$\nl
C37 & 2 & 511.9 & 454.3 & $24.47\pm 0.01$ & $23.56\pm 0.02$ & $50.46\pm 0.62$\nl
C38 & 2 & 210.9 & 130.1 & $25.74\pm 0.04$ & $24.57\pm 0.06$ & $40.30\pm 1.32$\nl
C39 & 2 & 213.3 &  96.5 & $25.65\pm 0.02$ & $24.49\pm 0.04$ & $39.61\pm 0.52$\nl
C40 & 3 & 324.0 & 178.1 & $25.68\pm 0.03$ & $24.65\pm 0.08$ & $28.51\pm 0.39$\nl
C41 & 3 & 196.2 & 181.9 & $25.88\pm 0.03$ & $24.98\pm 0.08$ & $17.94\pm 0.25$\nl
C42 & 3 & 230.4 & 187.3 & $24.62\pm 0.01$ & $23.66\pm 0.02$ & $55.90\pm 1.97$\nl
C43 & 3 & 562.4 & 190.4 & $25.82\pm 0.02$ & $24.92\pm 0.04$ & $33.63\pm 0.24$\nl
C44 & 3 & 455.5 & 194.4 & $25.65\pm 0.03$ & $24.61\pm 0.05$ & $36.06\pm 0.46$\nl
C45 & 3 & 257.3 & 213.7 & $24.67\pm 0.01$ & $23.94\pm 0.03$ & $44.54\pm 1.29$\nl
C46 & 3 & 335.4 & 214.0 & $25.72\pm 0.02$ & $24.97\pm 0.05$ & $25.10\pm 0.18$\nl
C47 & 3 & 479.3 & 228.3 & $25.50\pm 0.02$ & $24.67\pm 0.05$ & $38.46\pm 0.69$\nl
C48 & 3 & 614.2 & 231.1 & $24.55\pm 0.01$ & $23.71\pm 0.02$ & $49.25\pm 1.24$\nl
C49 & 3 & 381.5 & 280.7 & $24.98\pm 0.02$ & $23.82\pm 0.03$ & $38.47\pm 0.46$\nl
C50 & 3 & 412.2 & 294.8 & $25.41\pm 0.02$ & $24.28\pm 0.03$ & $46.23\pm 0.57$\nl
C51 & 3 & 281.8 & 301.0 & $26.09\pm 0.03$ & $25.46\pm 0.06$ & $18.13\pm 0.17$\nl
C52 & 3 & 761.0 & 323.7 & $26.10\pm 0.03$ & $25.36\pm 0.05$ & $17.32\pm 0.18$\nl
C53 & 3 & 233.2 & 347.9 & $25.50\pm 0.02$ & $24.56\pm 0.05$ & $33.49\pm 0.46$\nl
C54 & 3 & 503.7 & 422.6 & $26.43\pm 0.04$ & $25.74\pm 0.08$ & $16.04\pm 0.27$\nl
C55 & 3 & 408.4 & 492.1 & $25.52\pm 0.02$ & $24.70\pm 0.04$ & $20.55\pm 0.11$\nl
C56 & 3 & 306.6 & 786.6 & $25.10\pm 0.01$ & $24.32\pm 0.04$ & $30.23\pm 0.28$\nl
C57 & 3 & 402.2 & 790.9 & $25.16\pm 0.02$ & $24.20\pm 0.04$ & $45.37\pm 1.01$\nl
C58 & 3 & 567.6 & 732.9 & $25.62\pm 0.02$ & $24.87\pm 0.04$ & $20.75\pm 0.21$\nl
C59 & 3 & 454.1 & 238.2 & $25.98\pm 0.03$ & $24.95\pm 0.08$ & $22.20\pm 0.53$\nl
C60 & 3 & 357.0 & 282.0 & $25.04\pm 0.02$ & $24.37\pm 0.04$ & $33.82\pm 0.29$\nl
C61 & 3 & 292.1 & 619.2 & $25.63\pm 0.03$ & $24.80\pm 0.06$ & $23.90\pm 0.36$\nl
C62 & 3 & 120.9 & 428.7 & $24.47\pm 0.02$ & $23.44\pm 0.04$ & $52.38\pm 0.67$\nl
C63 & 3 & 101.1 & 438.2 & $25.18\pm 0.02$ & $24.51\pm 0.04$ & $33.03\pm 0.41$\nl
C64 & 3 & 724.1 & 450.1 & $26.01\pm 0.02$ & $25.28\pm 0.04$ & $17.63\pm 0.11$\nl
C65 & 3 & 110.2 & 195.1 & $25.22\pm 0.03$ & $24.11\pm 0.03$ & $48.19\pm 1.93$\nl
C66 & 3 & 547.8 & 661.5 & $25.47\pm 0.02$ & $24.80\pm 0.04$ & $21.53\pm 0.27$\nl
C67 & 3 & 452.3 & 242.2 & $26.02\pm 0.03$ & $24.82\pm 0.05$ & $39.27\pm 0.78$\nl
C68 & 4 & 309.2 & 137.0 & $26.32\pm 0.03$ & $25.57\pm 0.08$ & $18.57\pm 0.21$\nl
C69 & 4 & 196.6 & 148.0 & $25.89\pm 0.04$ & $24.85\pm 0.07$ & $21.61\pm 0.40$\nl
C70 & 4 & 172.6 & 275.4 & $26.13\pm 0.04$ & $25.35\pm 0.11$ & $25.94\pm 0.24$\nl
C71 & 4 & 278.9 & 351.3 & $26.08\pm 0.03$ & $25.18\pm 0.07$ & $23.02\pm 0.42$\nl
C72 & 4 & 169.8 & 450.0 & $26.23\pm 0.05$ & $25.40\pm 0.08$ & $16.60\pm 0.20$\nl
C73 & 4 & 710.3 & 105.4 & $25.51\pm 0.02$ & $24.45\pm 0.03$ & $36.82\pm 0.61$\nl
C74 & 4 & 246.2 & 121.9 & $25.35\pm 0.02$ & $24.19\pm 0.03$ & $59.97\pm 1.64$\nl
C75 & 4 & 637.2 & 150.7 & $25.21\pm 0.02$ & $24.10\pm 0.03$ & $57.88\pm 1.00$\nl
C76 & 4 & 145.2 & 163.0 & $26.43\pm 0.04$ & $25.33\pm 0.06$ & $27.32\pm 0.50$\nl
C77 & 4 & 337.7 & 236.1 & $25.43\pm 0.03$ & $24.79\pm 0.03$ & $22.94\pm 0.26$\nl
C78 & 4 & 637.4 & 252.9 & $26.48\pm 0.04$ & $25.68\pm 0.08$ & $17.23\pm 0.18$\nl
C79 & 4 & 211.9 & 334.9 & $26.28\pm 0.04$ & $25.20\pm 0.08$ & $20.13\pm 0.11$\nl
C80 & 4 & 159.4 & 358.9 & $25.45\pm 0.03$ & $24.63\pm 0.03$ & $33.73\pm 0.70$\nl
C81 & 4 & 300.2 & 369.1 & $25.87\pm 0.03$ & $25.12\pm 0.06$ & $21.37\pm 0.27$\nl
C82 & 4 & 717.8 & 408.7 & $26.37\pm 0.03$ & $25.43\pm 0.08$ & $20.02\pm 0.12$\nl
C83 & 4 & 618.4 & 413.1 & $25.65\pm 0.02$ & $24.74\pm 0.04$ & $24.07\pm 0.25$\nl
C84 & 4 & 492.4 & 679.0 & $25.67\pm 0.02$ & $24.61\pm 0.04$ & $22.69\pm 0.27$\nl
C85 & 4 & 360.2 & 658.5 & $25.50\pm 0.02$ & $24.61\pm 0.04$ & $25.16\pm 0.21$\nl
C86 & 4 & 391.5 & 645.4 & $25.17\pm 0.01$ & $24.36\pm 0.03$ & $30.30\pm 0.32$\nl
C87 & 4 & 101.9 & 644.0 & $26.12\pm 0.04$ & $25.21\pm 0.07$ & $18.12\pm 0.18$\nl
C88 & 4 & 427.0 & 636.2 & $25.42\pm 0.02$ & $24.63\pm 0.05$ & $26.88\pm 0.20$\nl
C89 & 4 & 401.2 & 613.1 & $25.31\pm 0.02$ & $24.47\pm 0.03$ & $51.18\pm 1.18$\nl
C90 & 4 & 375.2 & 619.5 & $26.20\pm 0.04$ & $25.18\pm 0.11$ & $20.59\pm 0.36$\nl
C91 & 4 & 597.0 & 462.4 & $26.16\pm 0.04$ & $25.53\pm 0.07$ & $16.31\pm 0.15$\nl
C92 & 4 & 138.7 & 463.2 & $26.07\pm 0.03$ & $25.10\pm 0.06$ & $25.01\pm 0.42$\nl
C93 & 4 & 138.6 & 180.7 & $25.33\pm 0.02$ & $24.49\pm 0.05$ & $35.64\pm 0.55$\nl
C94 & 4 & 643.6 & 333.3 & $25.51\pm 0.02$ & $24.57\pm 0.04$ & $33.34\pm 0.55$\nl
\enddata
\end{deluxetable}

\clearpage

\epsscale{1.0}
\plotone{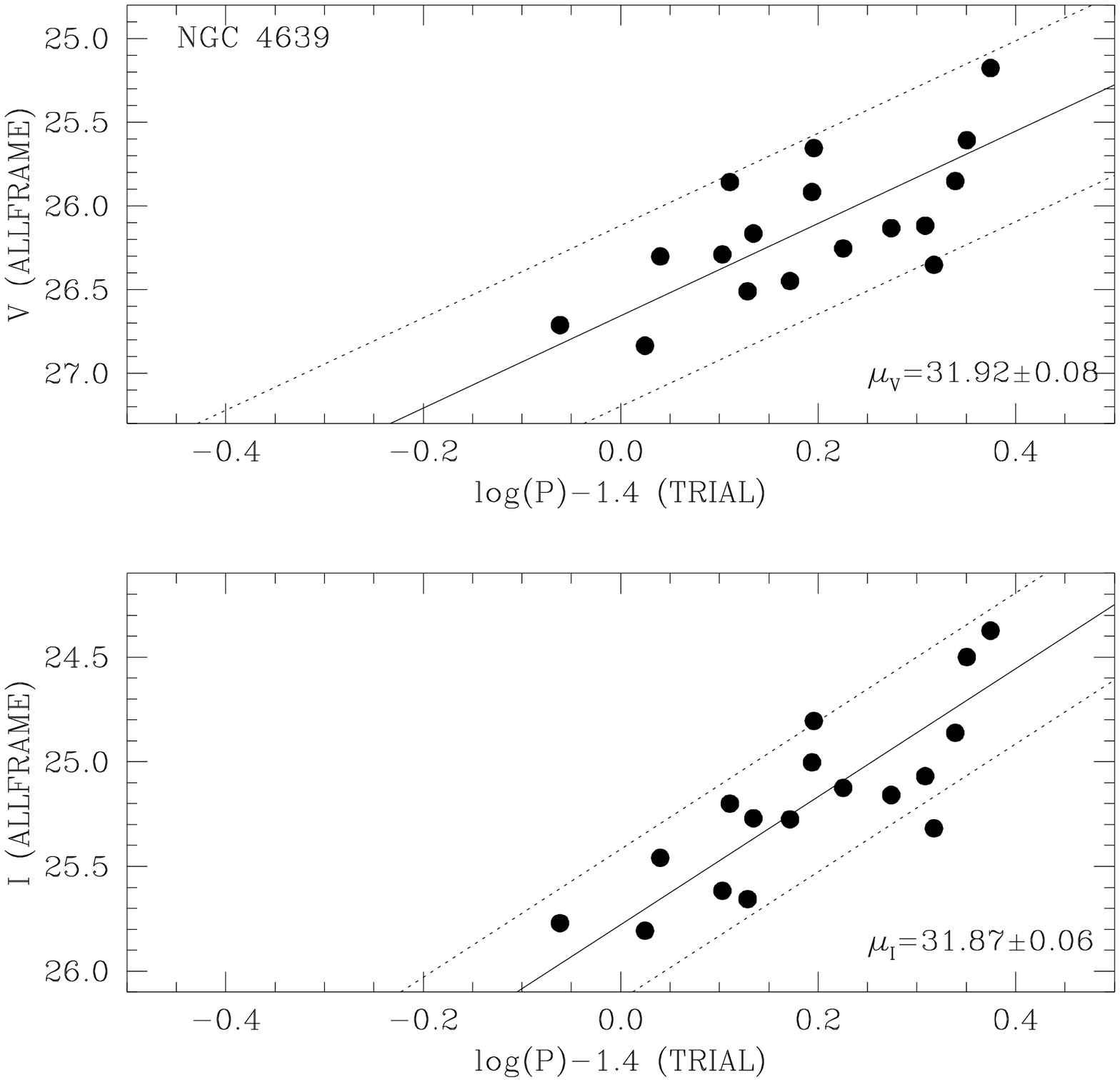}

\clearpage

\epsscale{1.0}
\plotone{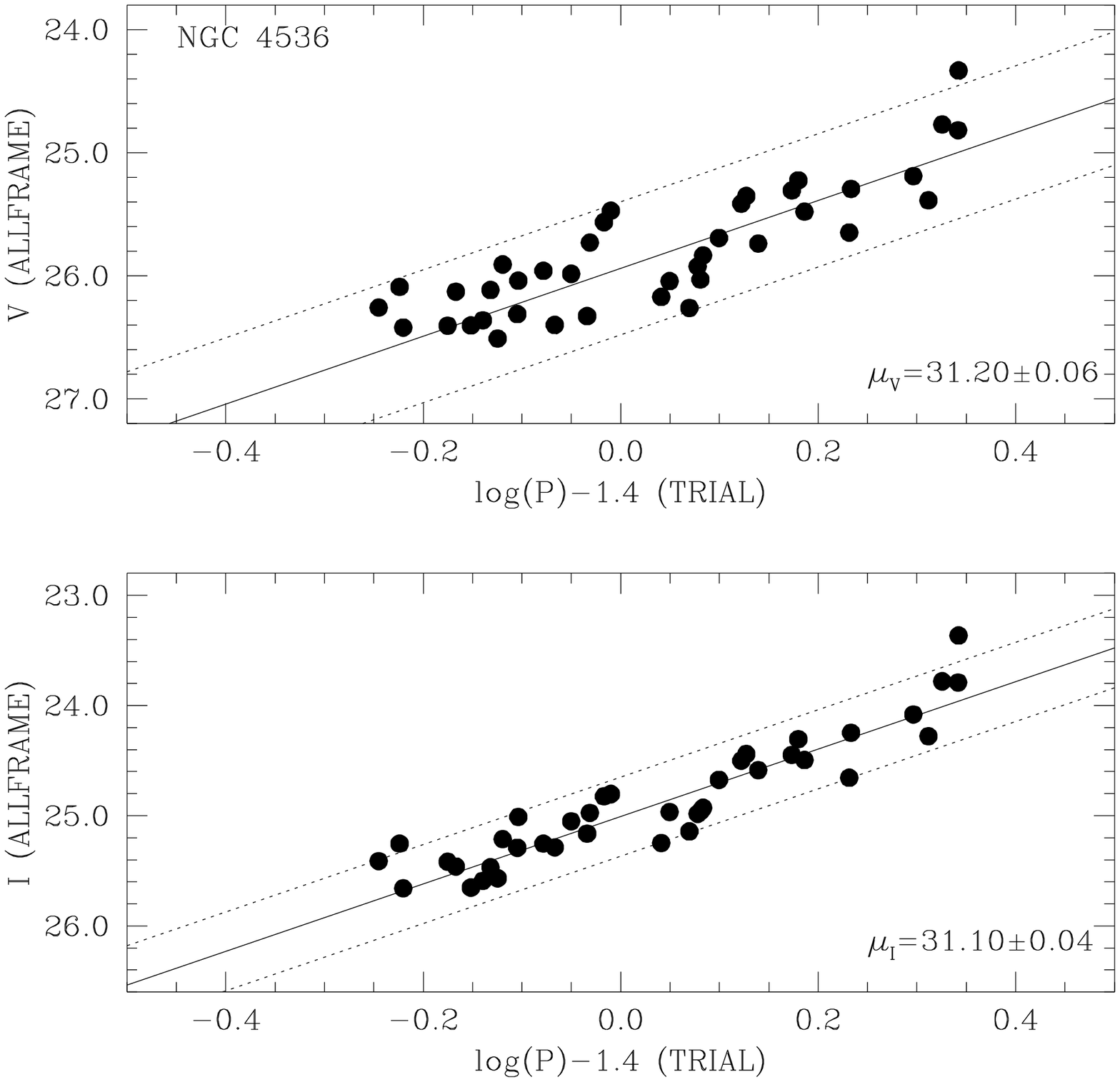}

\clearpage

\epsscale{1.0}
\plotone{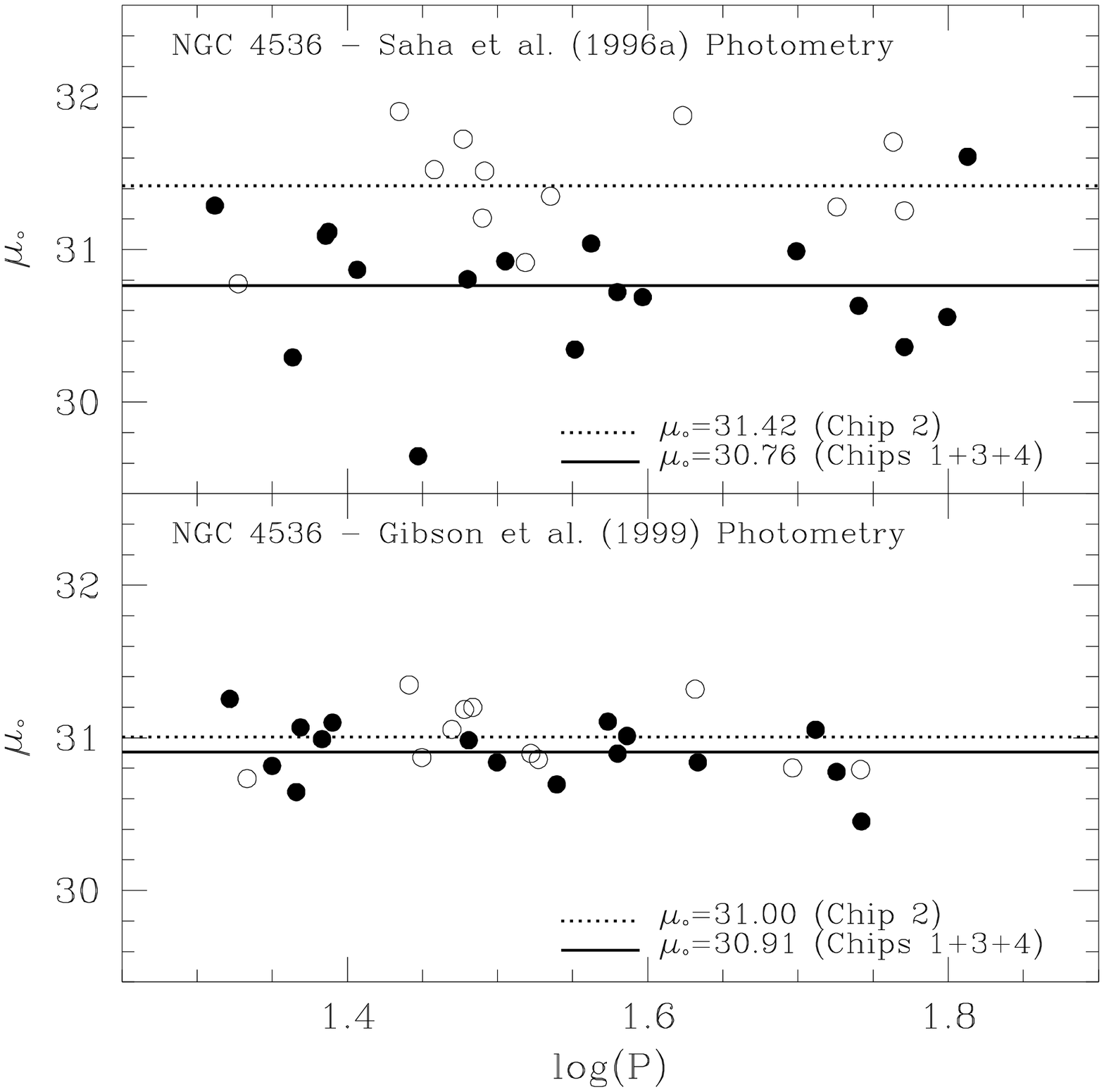}

\clearpage

\epsscale{1.0}
\plotone{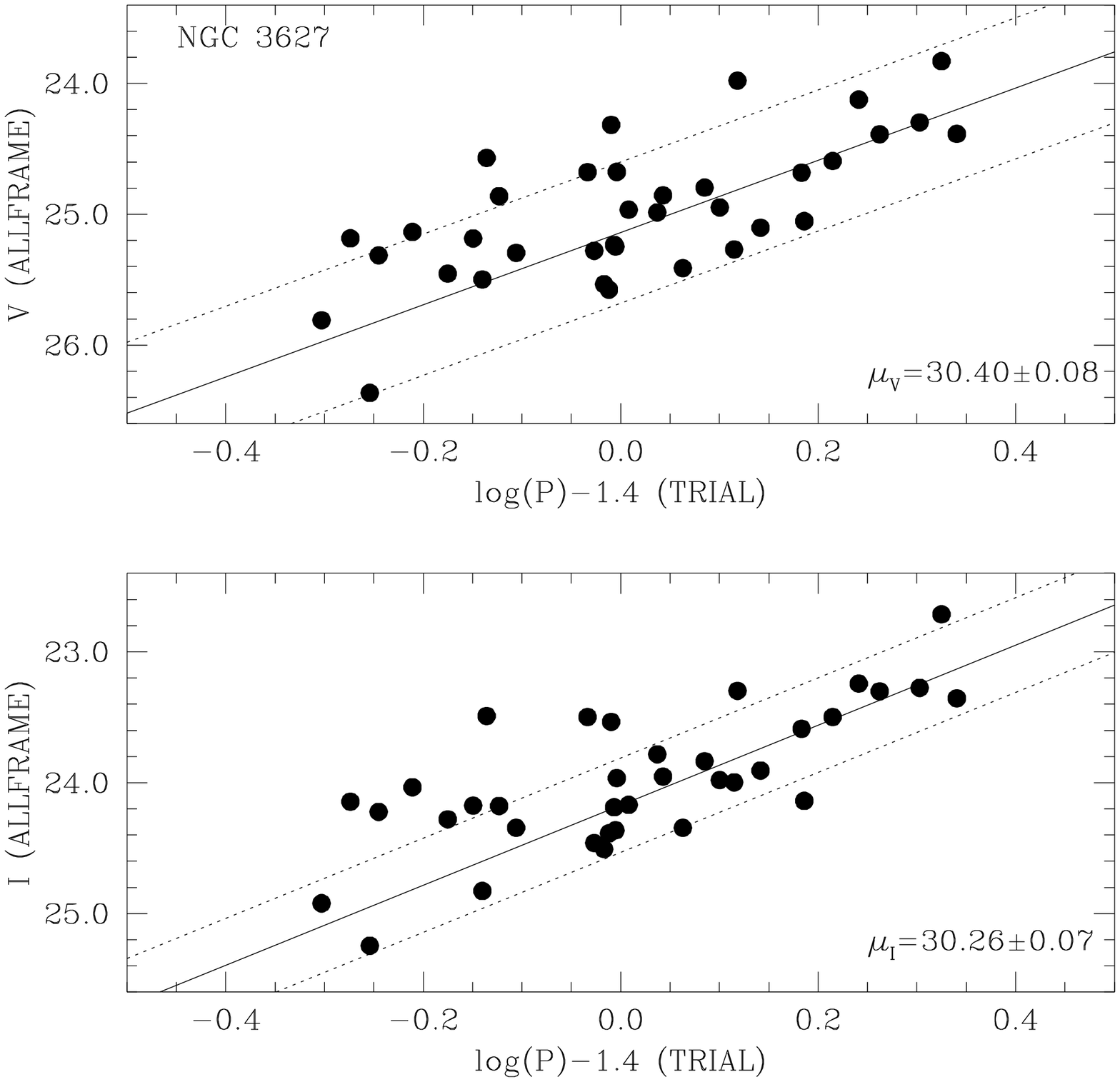}

\clearpage

\epsscale{1.0}
\plotone{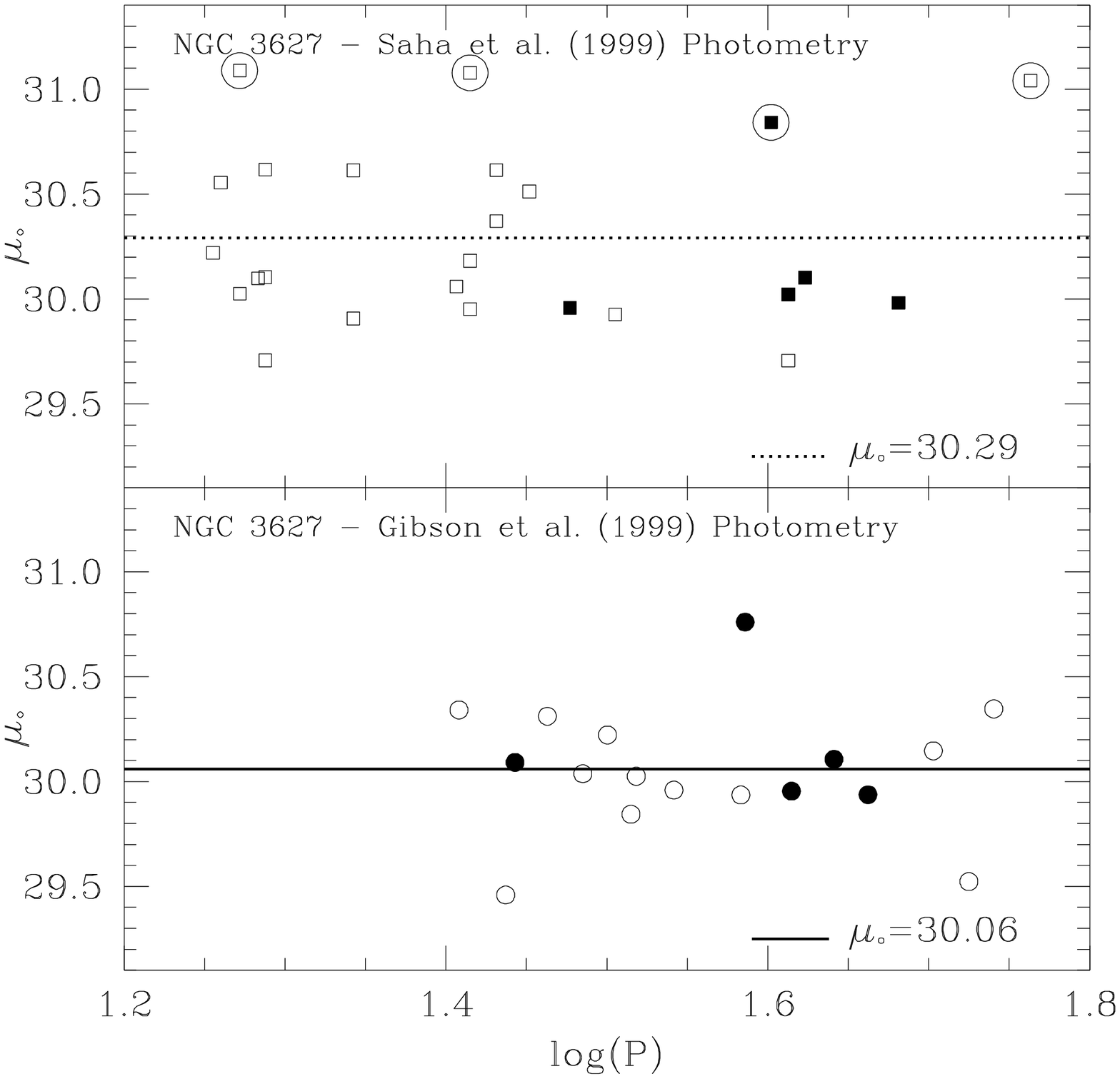}

\clearpage

\epsscale{1.0}
\plotone{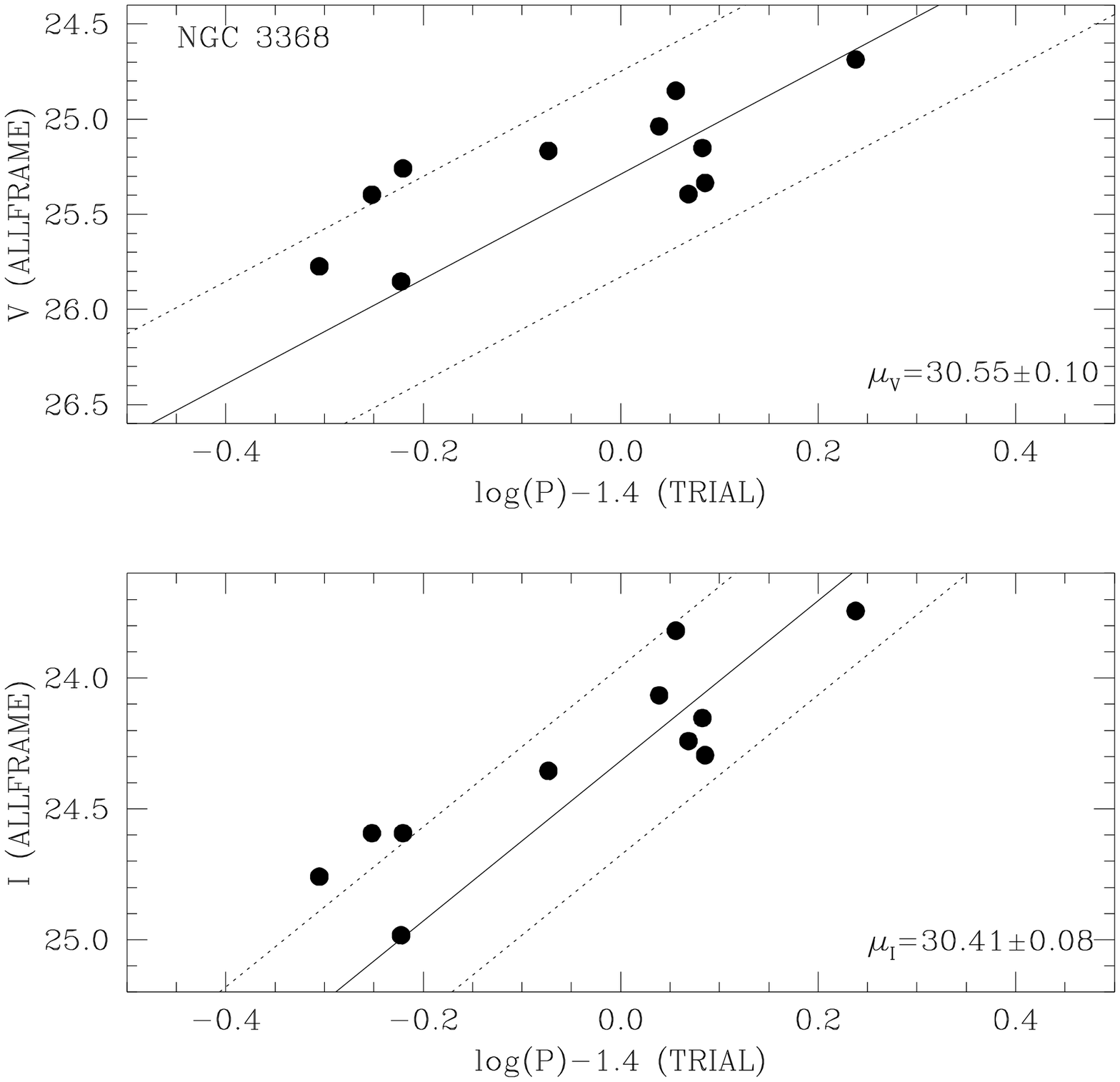}

\clearpage

\epsscale{1.0}
\plotone{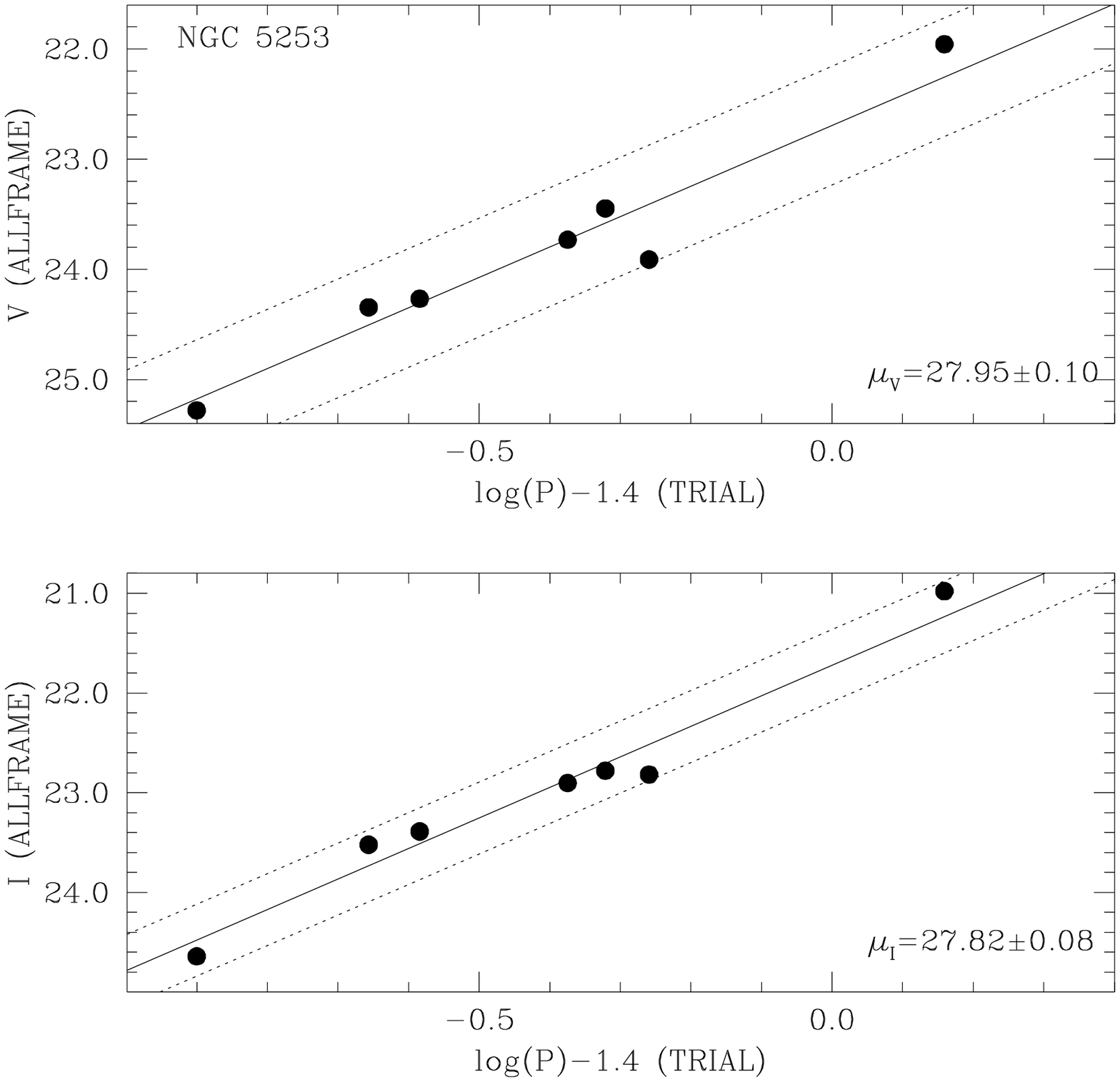}

\clearpage

\epsscale{1.0}
\plotone{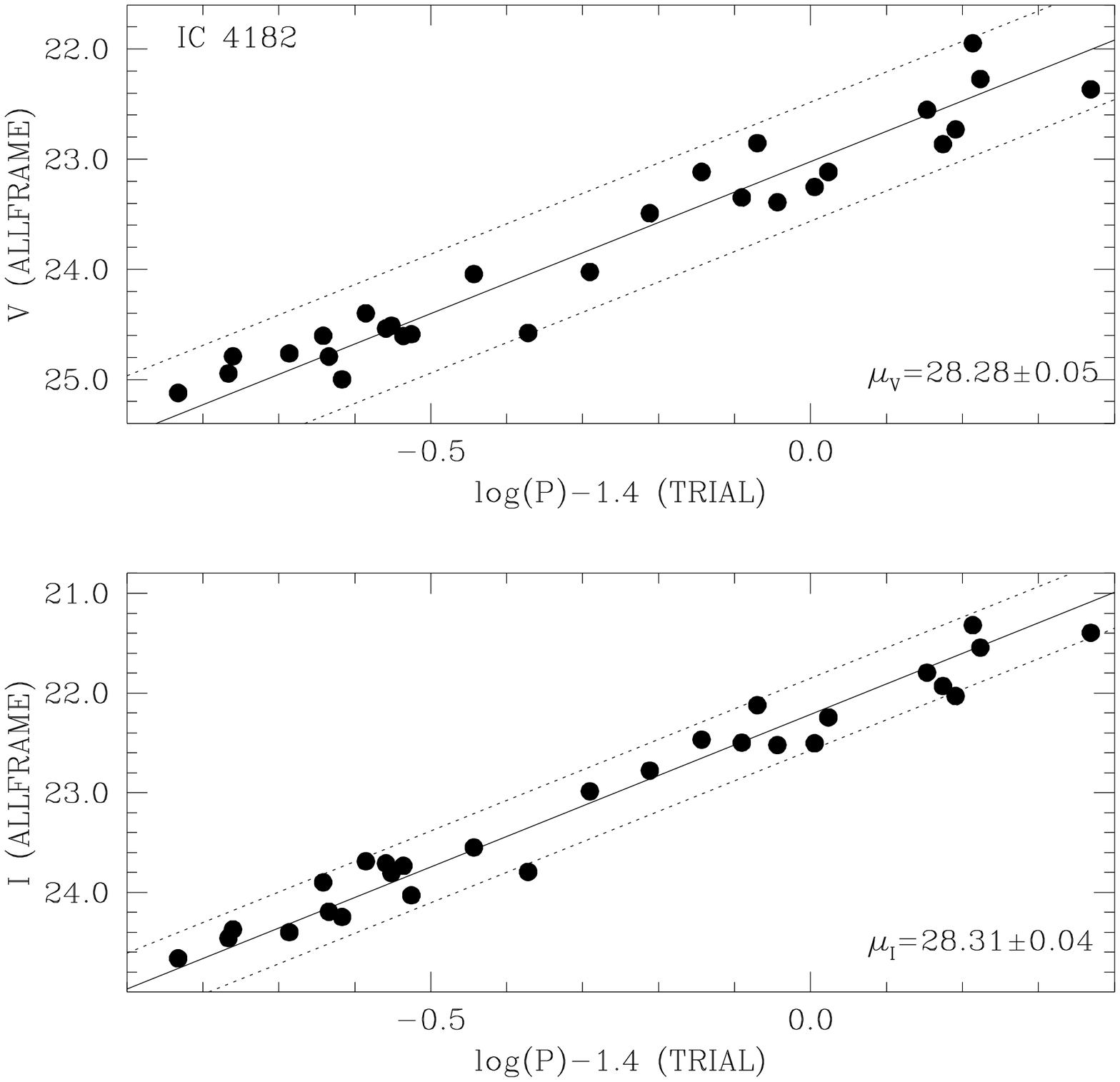}

\clearpage

\epsscale{1.0}
\plotone{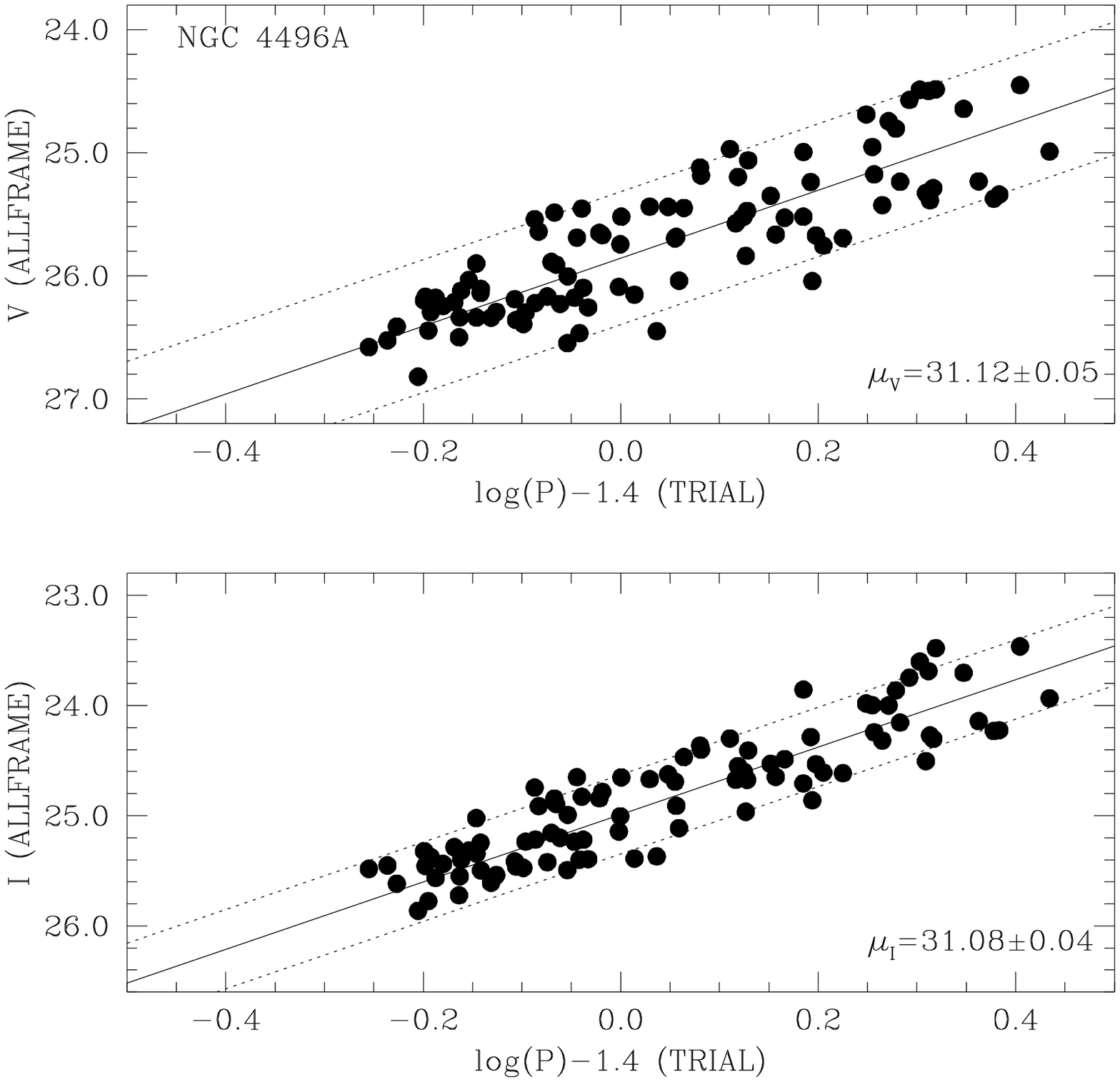}

\end{document}